\documentclass[12pt,a4paper]{article}

\usepackage[a4paper,left=1in,top=1.2in,bottom=1.4in]{geometry}
\usepackage{graphicx}
\usepackage{amsmath}
\usepackage{amssymb}
\usepackage{graphicx}
\usepackage{xspace}
\usepackage{psfrag}
\usepackage{hyperref}
\usepackage{varwidth}

\newcommand{\as}{\alpha_s}
\newcommand{\aEW}{\alpha_\text{\sc ew}}

\newcommand{\ee}{\ensuremath{e^+e^-}}

\newcommand{\nbar}{\ensuremath{\bar{n}}}
\newcommand{\LO}{\text{LO}\xspace}
\newcommand{\nLO}{\nbar \text{LO}\xspace}
\newcommand{\nNLO}{\nbar \text{NLO}\xspace}
\newcommand{\nnNLO}{\nbar\nbar \text{NLO}\xspace}
\newcommand{\nnLO}{\nbar\nbar \text{LO}\xspace}
\newcommand{\nNNLO}{\nbar \text{NNLO}\xspace}
\newcommand{\only}{\text{only}}
\newcommand{\GeV}{\ensuremath{\,\mathrm{GeV}}\xspace}
\newcommand{\TeV}{\ensuremath{\,\mathrm{TeV}}\xspace}

\newcommand{\jets}{\mathrm{jets}}
\newcommand{\tot}{\mathrm{tot}}
\newcommand{\jet}{\mathrm{jet}}
\newcommand{\reference}{\mathrm{ref}}
\newcommand{\order}[1]{\mathcal{O}\!\left(#1\right)}
\newcommand{\ie}{i.e.\ }

\newcommand{\muF}{\mu_\text{\sc f}}
\newcommand{\muR}{\mu_\text{\sc r}}
\newcommand{\LS}{\mathrm{LS}}
\newcommand{\nlojet}{NLOJet\texttt{++}\xspace}
\newcommand{\Z}{\mathrm{Z}}
\newcommand{\W}{\mathrm{W}}

\title{Giant QCD $K$-factors beyond NLO}
\author{Mathieu~Rubin, Gavin~P.~Salam and Sebastian~Sapeta\bigskip\\
  \normalsize LPTHE\\[-0.2em] 
  \normalsize UPMC Univ.\ Paris 6\\[-0.2em]
  \normalsize CNRS UMR 7589\\[-0.2em] 
  \normalsize Paris, France }
\date{}

\begin{document}
\maketitle
%

\begin{abstract}
  Hadronic observables in Z+jet events can be subject to large NLO
  corrections at TeV scales, with $K$-factors that even reach values
  of order 50 in some cases.
  We develop a method, LoopSim, by which approximate NNLO predictions
  can be obtained for such observables, supplementing NLO Z+jet and
  NLO Z+2-jet results with a unitarity-based approximation for missing
  higher loop terms.
  We first test the method against known NNLO results for Drell-Yan
  lepton $p_t$ spectra. 
  We then show our approximate NNLO results for the Z+jet observables.
  Finally we examine whether the LoopSim method can provide useful
  information even in cases without giant $K$-factors, with results
  for observables in dijet events that can be compared to early LHC
  data.
\end{abstract}

\newpage
\tableofcontents

\section{Introduction}
\label{sec:introduction}

At CERN's Large Hadron Collider (LHC), it is widely anticipated that
signals of new physics, for example supersymmetry, may manifest
themselves as large excesses of data compared to expected QCD and
electroweak backgrounds at high momentum
scales~\cite{Hinchliffe:1996iu,Abdullin:1998pm,Aad:2009wy,Ball:2007zza,Spiropulu:2008ug,Yamazaki:2008nm,Yamamoto:2007it}.
The estimation of these backgrounds will be one of the
elements in ascertaining the presence of any new physics from such
signals. Consequently, considerable effort is being invested across
the particle physics community in the development of methods to
understand and predict backgrounds (some of the issues involved are
described nicely in ref.~\cite{Mangano:2008ha}).

Given the QCD methods that are available today, some of the best
prospects for obtaining systematic, accurate predictions of
backgrounds involve next-to-leading order (NLO) QCD calculations.
By carrying out a systematic expansion in the strong coupling and
obtaining the first two terms (leading order (LO) and NLO) for a given
process, one often obtains predictions that are accurate to $10-20\%$.
The importance of NLO predictions in the LHC programme has motivated a
large calculational effort destined to extend the range of processes
known at NLO (for reviews, see
refs.~\cite{Bern:2008ef,Binoth:2010ra}).

While the majority of NLO calculations show some degree of convergence
relative to the LO results, several groups have commented in recent
years on the appearance of $K$ factors, ratios of NLO to LO
results, that grow dramatically towards high transverse momenta
\cite{Campbell:2007ws,Banfi:2007gu,Butterworth:2008iy,Bauer:2009km,Denner:2009gj,OleariTalk}
(similar behaviour is visible also in \cite{Cordero:2009kv,Binoth:2009wk}).
The problem generally occurs for hadronic observables (jet transverse
momenta, etc.) in processes that involve heavy vector bosons or heavy
quarks, at scales far above the boson or quark mass.

\begin{figure}
  \centering
  \includegraphics[width=0.95\textwidth]{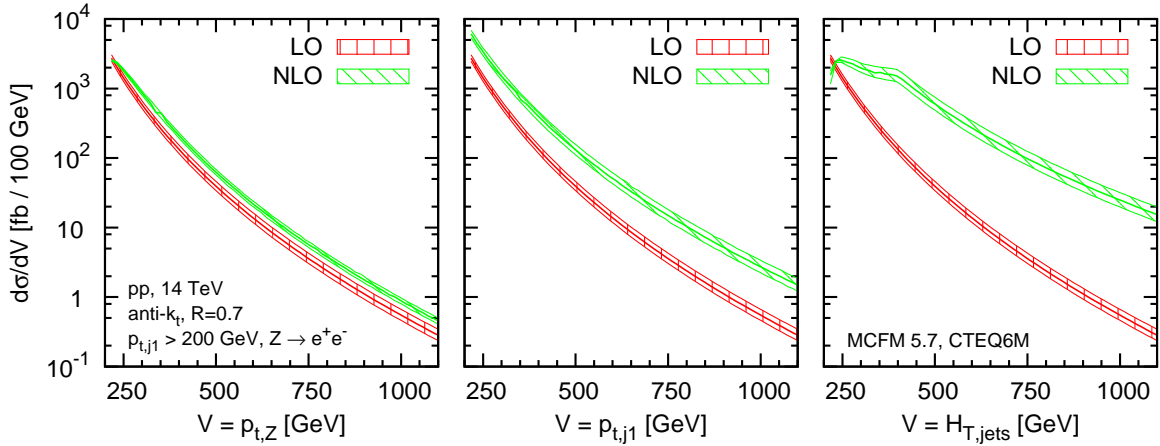}
  \caption{The LO and NLO distributions obtained with MCFM 5.7 \cite{mcfm}
    for three observables in
    Z+jet production: the Z transverse momentum (left), the $p_t$
    of the hardest jet (middle), and the scalar sum of the transverse momenta
    of all the jets, $H_{T,\jets}$ (right). 
    The  bands correspond to the uncertainty from a
    simultaneous variation of $\mu_{R}=\mu_F$ by a 
    factor of two either side of a default
    $\mu=\sqrt{\smash[b]{p_{t,j1}^2 + m_\Z^2}}$.
    The jet algorithm is anti-$k_t$ \cite{Cacciari:2008gp} with $R=0.7$
    and only events whose hardest jet passes a cut
    $p_t > 200\GeV$ are accepted.
    The cross sections include the branching ratio $\Z\to\ee$.  }
  \label{fig:LO_NLO_dist}
\end{figure}

Fig.~\ref{fig:LO_NLO_dist} illustrates this for the $pp\to\;$Z+jet
process at LHC ($14\TeV$) energies.
It shows the distributions of three observables that are non-zero for
configurations involving a Z-boson and one or more partons: the
transverse-momentum of the Z-boson ($p_{t,\Z}$), the
transverse-momentum of the highest-$p_t$ jet ($p_{t,j1}$) and the
effective mass (scalar sum of the transverse momenta) of all jets
($H_{T,\jets}$). At LO, all three distributions are identical.
At NLO, the $p_{t,\Z}$ observable is rather typical of a QCD
observable: its distribution has a NLO $K$-factor of about $1.5$,
fairly independently of $p_{t,\Z}$, and its scale dependence is
reduced with respect to LO.
The $p_{t,j1}$ distribution is more unusual: at high $p_t$ it has a
$K$-factor that grows noticeably with $p_{t,j1}$, reaching values of
about $4-6$, which is anomalously large for a QCD correction.
The $H_{T,\jets}$ observable is even more striking, with $K$-factors
approaching $100$.

\begin{figure}[t]
  \centering
  \begin{minipage}[b]{0.2\linewidth}
    \centering
    \scalebox{0.6}{\includegraphics{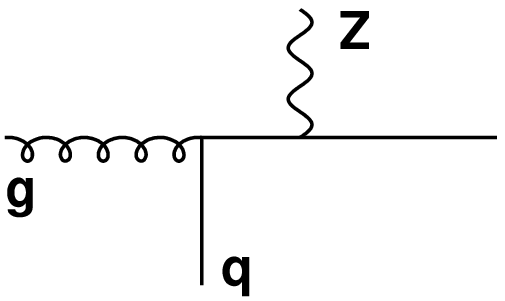}}\\
    (A)
  \end{minipage}
  \begin{minipage}[b]{0.2\linewidth}
    \centering
    \scalebox{0.6}{\includegraphics{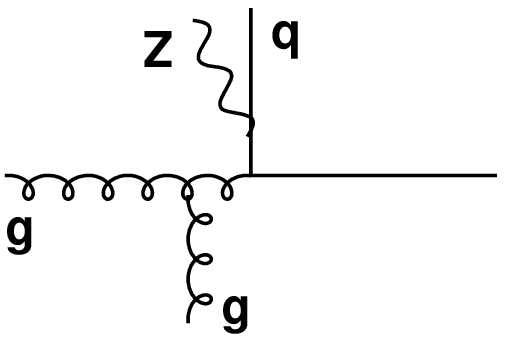}}\\
    (B)
  \end{minipage}
  \begin{minipage}[b]{0.2\linewidth}
    \centering
    \scalebox{0.6}{\includegraphics{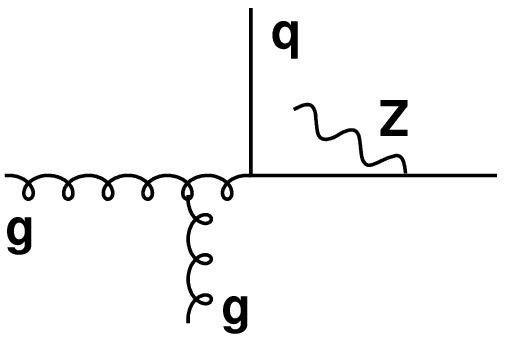}}\\
    (C)
  \end{minipage}
  \caption{A) a LO contribution to Z+jet production; B) and C) two
    contributions that are NLO corrections to Z+jet observables but
    whose topology is that of a dijet event with additional radiation
    of a soft or collinear Z-boson either from a final-state quark (B)
    or an initial-state one (C).}
  \label{fig:z+jet-diags}
\end{figure}

Given that fig.~\ref{fig:LO_NLO_dist} involves momentum scales where
$\as \sim 0.1$, one is driven to ask how it is that such ``giant''
$K$-factors can arise.
As touched on in \cite{Butterworth:2008iy}, and discussed in more
detail in \cite{Bauer:2009km,Denner:2009gj} for the $p_{t,j1}$ case,
the answer lies in the appearance of diagrams with new kinematic
topologies at NLO. 
This is illustrated in fig.~\ref{fig:z+jet-diags}: at LO the only
event topology (A) is that of a Z-boson recoiling against a quark or
gluon jet.
One type of NLO diagram involves gluon radiation from this basic
topology, giving modest corrections to all our observables.
However, there are also NLO diagrams (B,C) whose topology is that of a
dijet event, in which a soft or collinear Z-boson is radiated from
outgoing or incoming legs.
These diagrams do not contribute significantly to the $p_{t,\Z}$
distribution, because the Z-boson carries only a moderate fraction
of the total $p_t$.
However when examining $p_{t,j1}$, it is irrelevant whether the Z
boson is soft or not. Contributions B and C then lead to a result that
is of order $\as^2 \aEW \ln^2 p_{t,j1}/m_\Z$, where the
double logarithm comes from the integration over soft and collinear
divergences for Z emission.
The ratio of the NLO to LO results is therefore $\order{\as \ln^2
  p_{t,j1}/m_\Z}$,\footnote{This differs from double electroweak (EW)
  logarithms, which involve terms like $\alpha_\text{\sc ew} \ln^2
  p_t/m_\Z$ , and are usually much smaller. Examples do exist of
  ``giant'' EW effects when tagging flavour \cite{Ciafaloni:2006qu}. }
rather than just $\order{\as}$, hence the $K$-factor
that grows large with increasing $p_t$.\footnote{%
  Part of the enhancement at high $p_t$ also comes from the fact that
  one can have $qq \to qq $ scattering that emits a Z, whereas the
  $qq$ partonic channel does not contribute at LO.}
For the $H_{T,\jets}$ observable the enhancement is even bigger
because the dijet topology leads to $H_{T,\jets} \sim 2 p_{t,j1}$
instead of $H_{T,\jets} = p_{t,j1}$ at LO.

While it is reassuring that we can understand the physical origins of
the large $K$-factors in fig.~\ref{fig:LO_NLO_dist}, we are still left
with doubts as to the accuracy of the NLO Z+jet predictions for
$p_{t,j1}$ and $H_{T,\jets}$, since they are dominated by the LO
result for the Z+2-parton topologies.
One way forward would be to calculate the full NNLO corrections for
the Z+jet process.
However, while work is progressing on NNLO calculations of $2\to2$
processes with QCD final states (see e.g.\ \cite{Glover:2010im} and
references therein), results are not yet available; nor are they
likely to become available any time soon for some of the more complex
processes where giant $K$-factors have been observed (e.g.\ some
observables in $pp\to Wb\bar b$~\cite{Butterworth:2008iy,Cordero:2009kv}).
Alternatively one could simply try to avoid observables like
$p_{t,j1}$ and $H_{T,\jets}$ in inclusive event samples. For example,
with additional cuts on the vector-boson momentum or a second jet,
refs.~\cite{Butterworth:2008iy,Denner:2009gj} showed that the
$K$-factors are significantly reduced.
However, given the many analyses that are foreseen at the LHC, it is
likely that at least a few will end up probing regions where giant
$K$-factors are present.

To understand how else one might address the problem of giant $K$-factors,
one can observe that in our $\Z+\jet$ example, the bottleneck in
obtaining a NNLO prediction is the inclusion of the two-loop $2\to
Z+1\,\mathrm{parton}$ contributions and proper cancellation of all
infrared and collinear divergences.
Yet the two-loop (and squared one-loop) contribution will have the
topology of diagram A in fig.~\ref{fig:z+jet-diags} and should not be
responsible for the dominant part of the NNLO correction, which will
instead come from diagrams with the topology of B and C, with either
an extra QCD emission or a loop.
So if one includes tree-level $2\to Z+3$ and 1-loop $2\to Z+2$
diagrams (i.e.\ $\Z+2\;\jets$ at NLO) and supplements them with even a
crude approximation to the two-loop $2\to Z+1$ result, one that
suffices merely to cancel all divergences, then one should have a good
approximation to the full NNLO result (a related observation has been
exploited to obtain approximate NNLO results for high-$p_t$ J/$\psi$
production in \cite{Artoisenet:2008fc}).

The purpose of this article is to develop a general method for
obtaining such rough estimates of missing loop corrections. Our approach,
called LoopSim, will be based on unitarity.
After explaining how it works in section~\ref{sec:loopsim-section},
and outlining a secondary ``reference-observable'' approach for
control purposes, we will test the method by comparing its results to
full NNLO predictions for lepton-$p_t$ spectra in Drell-Yan production
in section~\ref{sec:DY-comparison}, apply it to our Z+jet observables
in section~\ref{sec:results} and finally, in section~\ref{sec:QCD},
examine whether it can be of use even in the absence of giant
$K$-factors, specifically for a number of dijet observables.

\section{The LoopSim method}
\label{sec:loopsim-section}

The main ingredient of the LoopSim method is a procedure for taking a
tree-level event with $n$ final state particles and supplementing it
with a series of events with $n-1$ particles (approximate 1-loop
events), $n-2$ particles (approximate 2-loop events), etc., such that
the sum of the weights of the full set of events is zero.  This
``unitarity'' property will ensure that all the soft and collinear
divergences of the tree-level matrix elements will cancel against
identical divergences in the simulated loop contributions.

\begin{figure}[t]
  \centering
  \includegraphics[width=0.8\textwidth]{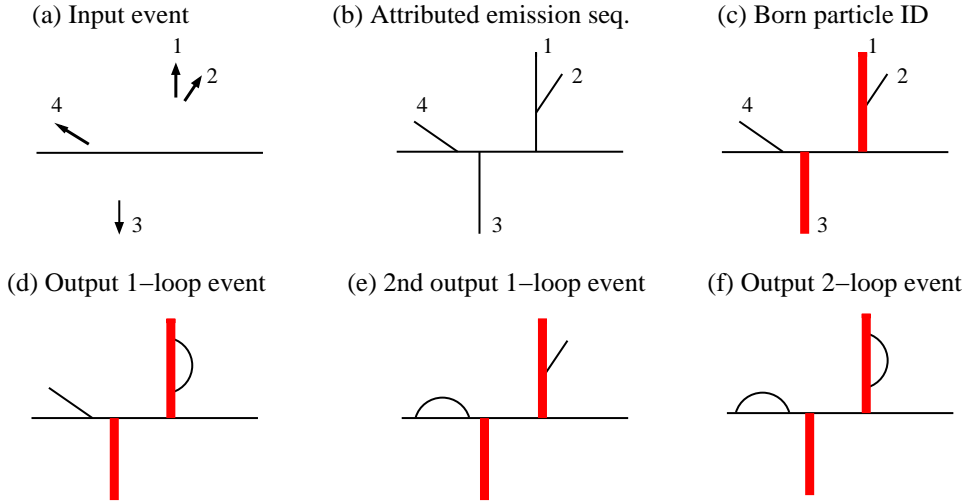}
  \caption{Sketch of the LoopSim procedure as applied to a tree-level
    event (a) with 4 outgoing particles (numbered) and the beam
    (horizontal line); diagram (b) shows the attribution of the
    emission sequence, (c) the identification of the Born particles
    (thick red lines), and (d)-(f) the resulting ``looped'' diagrams.
    These diagrams are relevant in approximating
    next-to-next-to-leading corrections to a process whose LO
    contribution has a $2\to 2$ structure.  }
  \label{fig:loopsim-diag}
\end{figure}

An outline of the procedure is given in fig.~\ref{fig:loopsim-diag}.
Given a tree-level input event (a), the first step is to interpret it
as a sequence of emissions (as if it had been produced by a parton
shower), so that for example (diagram b) one can view particle 2 as
having been emitted from particle 1, and particle 4 as emitted from
the beam.
The attribution of an emission sequence can be performed with the help
of a suitable sequential-recombination jet algorithm and will be most
meaningful in the limit that emissions are strongly ordered in angle
and energy.
The next stage is to decide which particles reflect the underlying
hard structure of the event. If the event structure at the lowest
possible order is that of a $2\to 2$ scattering, then one should
identify two outgoing ``Born'' particles. 
The Born particles will remain present in all the approximate ``loop''
events that are generated.
They are represented as thick red lines in diagram (c).
Again this step is most meaningful when all non-Born emissions are
soft and collinear.

One then generates a set of simulated ``1-loop'' events by finding all
ways of recombining one emitted particle with its emitter, diagrams
(d,e). 
Each such ``1-loop'' event comes with a relative weight of $-1$ compared
to the tree-level diagram.
Similarly the set of simulated ``2-loop'' events is obtained by
finding all ways of recombining two emitted particles with their
emitter(s) (diagram f), each with relative weight $+1$; and so forth
down to events where only Born particles remain (in
fig.~\ref{fig:loopsim-diag} this is already reached at the two-loop
level).
Note that the loop-diagrams drawn in fig.~\ref{fig:loopsim-diag} are
not intended to represent the actual Feynman diagrams that would be
relevant at 1 and 2-loop level. Instead they indicate the way in which
we have approximated the loop divergences, as the unitarising
counterparts of the divergences that appear for each emission in the
soft and collinear limits.

Given the above procedure for unitarising tree-level events, we shall
see that it is then straightforward to extend it to event sets that
also include exact loop diagrams.

\subsection{The tree-level pure glue case}
\label{sec:loopsim-pure-glue}

We start by examining the LoopSim procedure in the simple case of
purely gluonic tree-level events. This will suffice to introduce most
of the relevant concepts.
Section \ref{sec:loopsim-flavour} will then discuss some of the
additional issues that arise for events with quarks and vector bosons,
while the handling of events sets that include exact loop diagrams
will be left to section~\ref{sec:loopsim-NLO}.

It is helpful to introduce some notation: 
Firstly, $b$ is the number of final-state particles present in the
lowest relevant order (\ie the number of final-state ``Born''
particles). For instance $b=2$ if considering higher-order corrections
to dijet events, as in fig.~\ref{fig:loopsim-diag}.
$E_n$ represents a generic event with $n$ final state particles. So
the starting event of fig.~\ref{fig:loopsim-diag} would be labelled
$E_4$.
Finally, $U_l^b$ will be an operator that acts on an event $E_n$ and
returns all the events at $l$ loops obtained from $E_n$ using the
LoopSim method. For instance, fig.~\ref{fig:loopsim-diag}d,e
represents the action of $U_{l=1}^{b=2}$ on the input $E_4$ event (a).

The central part of the LoopSim method involves the construction of
the operator $U_l^{b}$ acting on $E_n$ for all $l=0\mathellipsis n-b$
($l\le n-b$ because the number of real final state particles cannot be
smaller than that of the lowest order event).
%

\subsubsection{Attribution of structure to events}
\label{sec:assignment-structure}

Recall that the primary function of the LoopSim method is to cancel
the divergences that appear in the soft and collinear limits.
In these limits, events can be interpreted as stemming from a sequence
of probabilistic (parton-shower) type $1\to 2$ splittings of some
original hard Born particles.
The knowledge of the splitting structure will help us generate loop
events to cancel the divergences.

The attribution of a branching sequence is most easily performed using
a sequential recombination jet algorithm (and is inspired by the
CKKW matching procedure~\cite{Catani:2001cc}).
We will use the Cambridge/Aachen (C/A) algorithm
\cite{Dokshitzer:1997in,Wobisch:1998wt}, which has important
advantages over the $k_t$ algorithm when dealing with nested collinear
divergences (avoiding ``junk'' jets
\cite{Dokshitzer:1997in}).\footnote{All jet clustering in this article
  is carried out using FastJet~\cite{Cacciari:2005hq,FastJet}.}

As a first step, to each of the $i = 1\ldots n$ particles in the event
$E_n$, we assign a unique ``identity'' index $I_i \equiv i$.

We then run the Cambridge/Aachen (C/A) algorithm on the event.
It repeatedly clusters the pair of particles that are closest in angle,
i.e.\ with smallest $d_{ij}=\Delta R_{ij}^2/R_{\LS}^2$ where $\Delta
R_{ij}^2=(y_i-y_j)^2+(\phi_i-\phi_j)^2$ is the usual squared angular
distance in the $(y,\phi)$ plane, and $R_\LS$ is a free parameter,
the LoopSim radius.
The C/A algorithm continues until all the $d_{ij} > 1$, at which point
the remaining particles are deemed to cluster with the beam.

An $ij \to k$ clustering in the C/A algorithm can be reinterpreted as
a $k \to ij$ splitting. The C/A algorithm does not distinguish in any
way between $i$ and $j$. 
However, in the soft limit, say $p_{tj} \ll p_{ti}$, rather than
viewing $k$ as splitting to $i$ and $j$, it is a better reflection of
the divergent structure of the amplitude to view $k$ as having emitted
a soft gluon $j$.
Then $i$ is nothing other than particle $k$ with some small fraction
of its energy removed.
To account for this, in an $ij \to k$ clustering, if
$p_{tj}<p_{ti}$, then we declare that the
``identity'' $I_k$ of particle $k$ should be the same as that of
particle $i$, $I_k = I_i$. Also we record $I_i$ as being a ``secondary
emitter'' and remember that the object with identity $I_j$ has been
emitted from the object with identity $I_i$.
(Exchange $i \leftrightarrow j$ if $p_{ti} < p_{tj}$).
This is represented in fig.~\ref{fig:loopsim-diag}b by the fact that
particle $1$ is a straight line, off which particle $2$ has been
emitted; the identity of the $1+2$ combination is $I_{1+2} \equiv
I_1 \equiv 1$.
For an $iB$ clustering, we record $I_i$ as having been emitted from
the beam.

The next step in attributing structure to the event is to decide which
event particles should be viewed as the Born particles, i.e.\ which
particles are responsible for the hard structure in the event.
Inspired by the original formulation of the Cambridge algorithm
\cite{Dokshitzer:1997in}, for every $ij\to k$ recombination we assign
a $k_t$ algorithm type hardness measure
$h_{ij}=\min(p_{ti}^2,p_{tj}^2)\Delta
R_{ij}^2/R_{\LS}^2$~\cite{Kt,Kt-EllisSoper}.\footnote{In the results
  shown later, we actually 
  used $h_{ij}=\min(p_{ti}^2,p_{tj}^2)\Delta R_{ij}^2$, which
  however is identical for our default choice of $R_\LS=1$.}
For every beam recombination, we assign a
hardness $h_{iB} = p_{ti}^2$.

We then work through the recombinations in order of decreasing
hardness.
For an $ij\to k$ recombination (or $k\to ij$ splitting), assuming $i$
is harder than $j$, we mark $I_k\equiv I_i$ as a Born particle. If
fewer than $b$ particles have already been marked as Born particle, we
also mark $I_j$ as a Born particle.
For an $iB$ recombination, we mark $I_i$ as a Born particle.
This is repeated until $b$ particles have been marked as Born (a
particle may be marked more than once; in such a case its marking
counts only once).
As an example, in fig.~\ref{fig:loopsim-diag}, the hardest
recombination will be between particle $3$ and the beam, so particle
$3$ is marked as a Born particle.
The next hardest recombination will that of $(1+2)$ with the beam. 
Therefore we mark $I_{(1+2)} = I_1 = 1$ as a Born particle.
This exhausts the number ($b=2$) of Born particles that need to be
marked.

At the end of the above procedure, every particle will have been
marked as emitted either from the beam or from another particle, and
some particles will also have been marked as secondary emitters and/or Born
particles.
Thus in figure~\ref{fig:loopsim-diag}, particle $1$ is labelled as
having been emitted from the beam, it is a secondary emitter and a
Born particle; particle $2$ is labelled as having been emitted from
particle $1$; particle $3$ is a Born particle, emitted from the beam;
and particle $4$ was emitted from the beam.
The structure that we attribute is of course physically unambiguous
only in the presence of strong ordering of emission angles and
energies. However, as we shall argue in
section~\ref{sec:loopsim-expected-accuracy}, the mistakes
that we make for non-ordered configurations should have a small impact
for observables with giant $K$-factors.

\subsubsection{Constructing virtual (loop) events}
\label{sec:looping-particles}

Once every particle is labelled in an event $E_n$, one can compute the
result of $U_{l}^{b}(E_n)$, which is a set of events $E_{n-l}$. For an
event $E_n$ with respectively $b$ Born particles and $n_s$ non-Born
secondary emitters, we define
\begin{equation}
  v\equiv n-(b+n_s)\,,
\end{equation}
to be the maximum number of particles that will be allowed to become
virtual in a given event. 
It is obvious that Born particles will not become virtual.
Additionally, secondary emitters will also not become virtual. To
understand why, consider the event
\begin{equation}
  \begin{varwidth}{0.3\linewidth}\centering
    \includegraphics[scale=0.4]{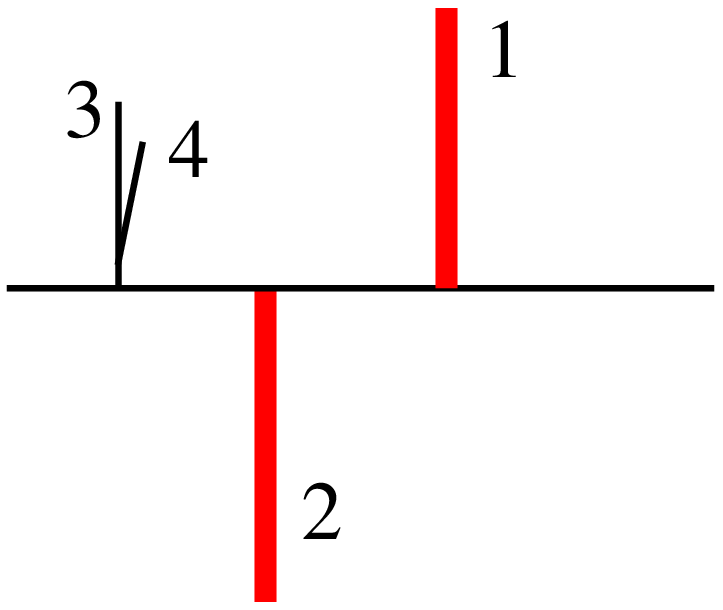}
  \end{varwidth}
\end{equation}
in which particle $3$ is a secondary emitter, since it emitted $4$. 
There is a divergence for $4$ to be collinear to $3$ only if $3$ is a
final-state particle. If instead $3$ is made virtual, then the
divergence for emitting $4$ no longer exists (there is no divergence for
emission from internal lines in a diagram), so that the weight for the diagram
in which $3$ is virtual would be the weight of the tree-level diagram times a
small coefficient $\varepsilon \ll 1$.
This simplest way of accounting for this is to approximate
$\varepsilon = 0$ and thus not generate events in which secondary
emitters are made virtual (a more detailed discussion is given in
appendix~\ref{sec:secondary-emitters}).

Having understood which particles can be made virtual, the operator
$U_{l}^b$, when applied on an event $E_n$, generates all the
$\binom{v}{l}$ diagrams in which $l$ particles become virtual.
For the virtual events to cancel the infrared and collinear
divergences that appear in the tree-level diagram, we need an
infrared and collinear (IRC) safe procedure to make particles
virtual.
For instance, the divergent weight of an event with two collinear
partons $i$ and $j$ has to be cancelled by that of corresponding
virtual event ($j$ makes a loop over $i$) when computing the
distribution of any IRC safe observable; and two collinear partons, if
not virtualised, have to remain collinear when another particle
becomes virtual. 

There are two ways for a particle $j$ to make a loop:
\begin{itemize}
\item If it is labelled as clustering with particle $i$, then one has
  to spread the momentum of particle $j$ over $i$ and all the
  particles that are labelled as clustering with it but which were
  emitted after $j$ according to the C/A clustering sequence (i.e.\ at
  smaller angle). The exact procedure is explained in detail in
  appendix~\ref{app:recoil_procedure}, and is designed to ensure that
  the recombination maintains any collinearity properties of non-looped particles
  and is invariant under longitudinal boosts. When $j$ is the only
  particle that clusters with $i$, then the procedure becomes
  equivalent to adding the momenta of particles $i$ and $j$,
  $p_k=p_i+p_j$, and then rescaling the momentum $p_k$ such that its
  mass is set to $0$, while leaving its transverse components $p_x$,
  $p_y$ and its rapidity unchanged.
\item If particle $j$ is labelled as clustering with the beam, then
  when it is ``looped'' it is simply removed from the event. 
  Note that looping particles with the beam is less trivial than it
  may seem at first sight, because of an interplay with factorisation
  and the PDFs. Nevertheless it can be shown,
  appendix~\ref{sec:incoming-partons}, that for particle types that
  are included in the PDFs it does make sense to loop them.
  A $p_t$ imbalance will result from the looping of particles with the
  beam, and so after all loops have been made, we apply a transverse
  boosts to all remaining event particles, conserving their
  rapidities, so as to bring the total transverse momentum to zero
  (again, see appendix~\ref{app:recoil_procedure}).
\end{itemize}
There is some arbitrariness to our procedures for producing physical
kinematics in the looped events. One avenue for future work would be
to examine the impact of making different choices.

The operator $U^b_l$ has the following properties
\begin{equation}
    \label{eq:Uprop}
    U_{0}^b =  1\!\!1\,, \qquad \qquad
    U_{l}^b(E_n)=0 \quad \text{ if }  l>v\,.
\end{equation}
If $w_n$ is the weight of event $E_n$, then each of the
events generated by the $U_{l}^b(E_n)$ operator has a weight
\begin{equation}
  w_{n-l} = (-1)^lw_n\,.
\end{equation}
Once all the $U_{l}^b(E_n)$ have been calculated for $l=0\mathellipsis
n-b$, one has to combine them in order to subtract all the soft and
collinear divergences that appear in the calculation of $E_n$ and the
virtual diagrams generated from it. This is done by the operator
$U_{\forall}^b$, which is defined as
\begin{equation}
  U_{\forall}^b \equiv \sum_{l=0}^{v}U_{l}^b\,.
\end{equation}
It generates all the necessary looped configurations that have the same order
in $\as$ as the original tree-level diagram. It is straightforward to
see that the total weight of the diagrams obtained from the
$U_{\forall}^b$ operator is $0$. Indeed, if we apply it to an event
$E_n$ whose maximum number of virtual particles is $v$, we get
\begin{equation}
  w_n\sum_{l=0}^{v}(-1)^l\binom{v}{l} = 0\,,\label{total_weight_is_0}
\end{equation}
for $v>0$.

We note that the above procedure for approximating loop diagrams does
not generate the finite terms needed to cancel the scale-dependence of
lower-order diagrams.
While it would be straightforward to include such terms, we believe that in
the absence of full loop calculations, not including them helps ensure
that the standard procedure of variation of renormalisation and
factorisation scales is more likely to provide some form of reasonable
estimate of the uncertainties on our results.

\subsubsection{Some examples}
\label{sec:loopsim-glue-examples}

In order to illustrate the action of the operator $U_{l}^b$, we give
below some simple examples in the pure glue case. In each of these
examples, only the Born particles are labelled with numbers 

{\allowdisplaybreaks 
  \begin{subequations}
    \label{eq:U-b-l-series}
    \begin{eqnarray}
      \label{U_4_1_2}
      U_{l=1}^{b=2} \left(
        \begin{varwidth}{\textwidth}
          \includegraphics[scale=0.35]{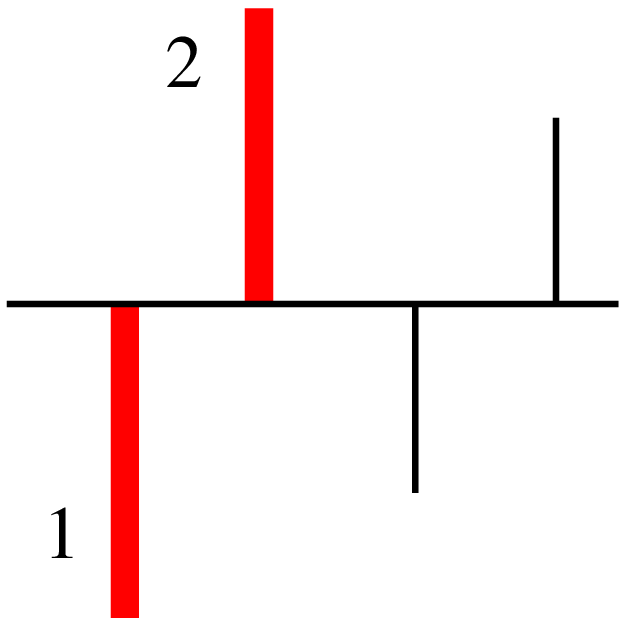}
        \end{varwidth}
      \right) & = &
      -\; 
      \begin{varwidth}{\textwidth}
        \includegraphics[scale=0.35]{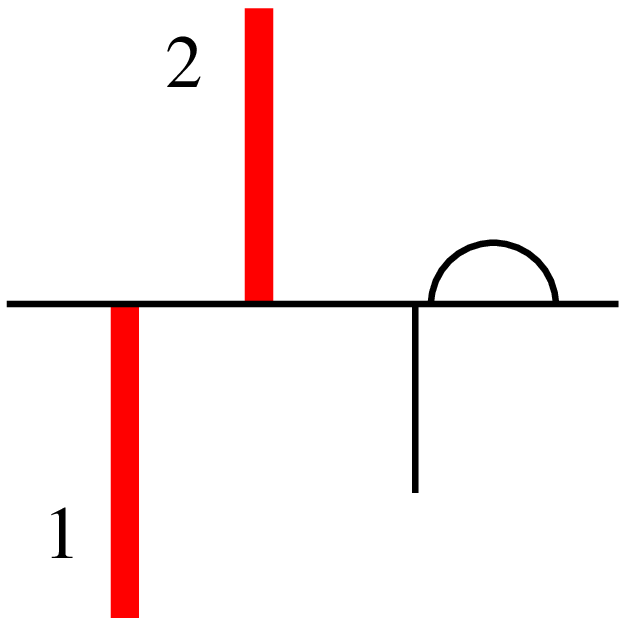}
      \end{varwidth}
      \;- \;
      \begin{varwidth}{\textwidth}
        \includegraphics[scale=0.35]{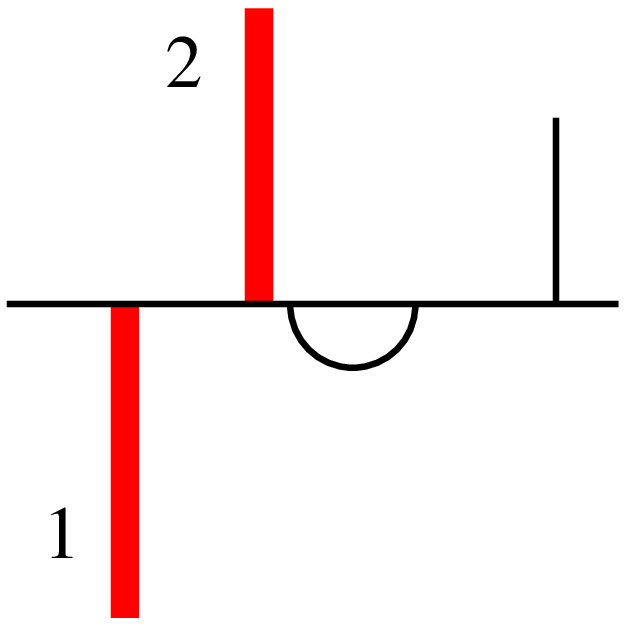}
      \end{varwidth}
      \,,  \\
      \label{U_4_2_2}
      U_{l=2}^{b=2} \left(
        \begin{varwidth}{\textwidth}
          \includegraphics[scale=0.35]{figs/ex_b201.eps}
        \end{varwidth}
      \right) & =  &
      \begin{varwidth}{\textwidth}
        \includegraphics[scale=0.35]{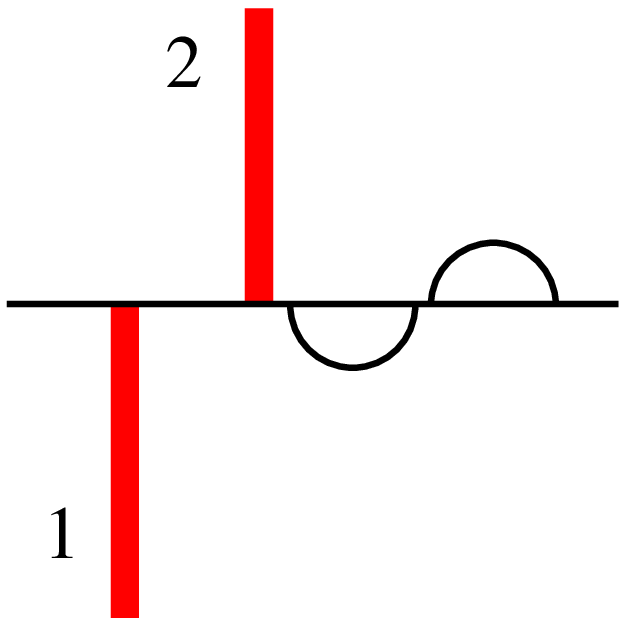}
      \end{varwidth}
      \,, \\
      \label{U_5_2_2_1}
      U_{l=2}^{b=2} \left(
        \begin{varwidth}{\textwidth}
          \includegraphics[scale=0.35]{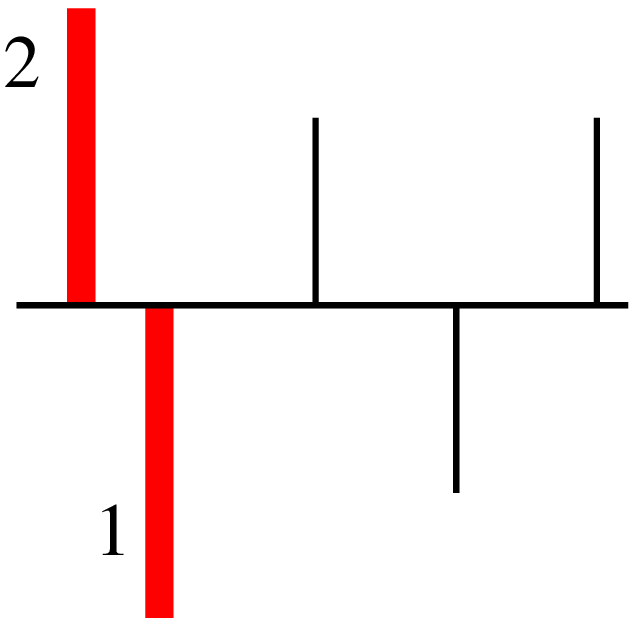}
        \end{varwidth}
      \right) & = &
      \begin{varwidth}{\textwidth}
        \includegraphics[scale=0.35]{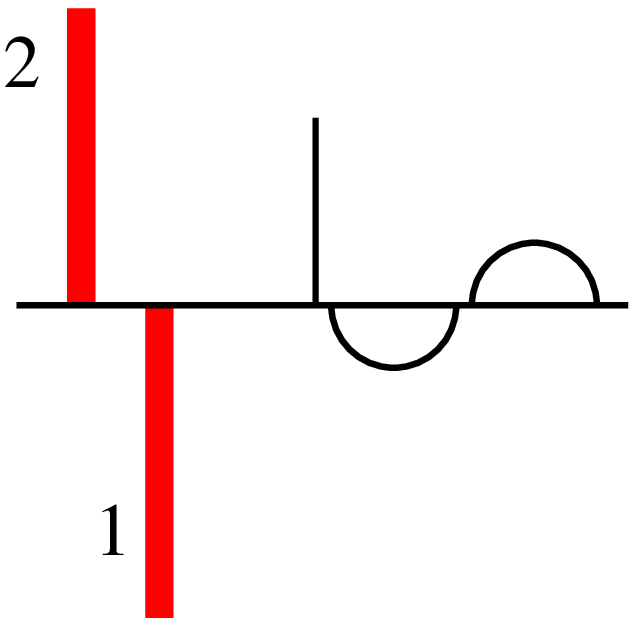}
      \end{varwidth}
      \;+ \;
      \begin{varwidth}{\textwidth}
        \includegraphics[scale=0.35]{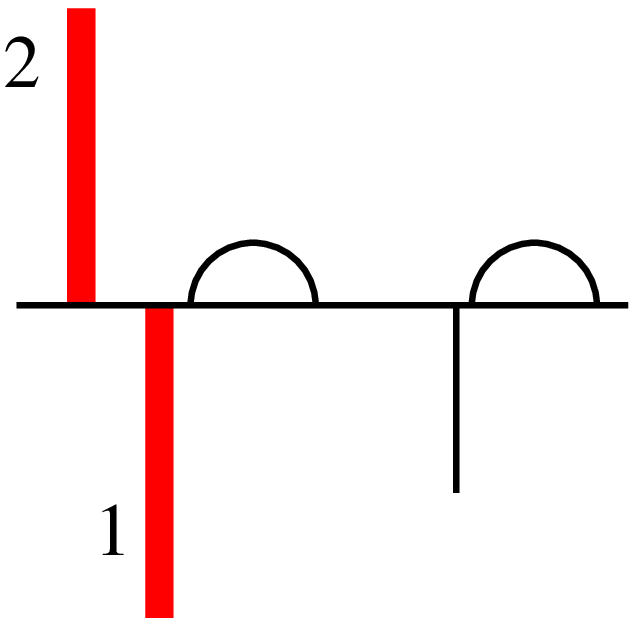}
      \end{varwidth}
      \;+ \;
      \begin{varwidth}{\textwidth}
        \includegraphics[scale=0.35]{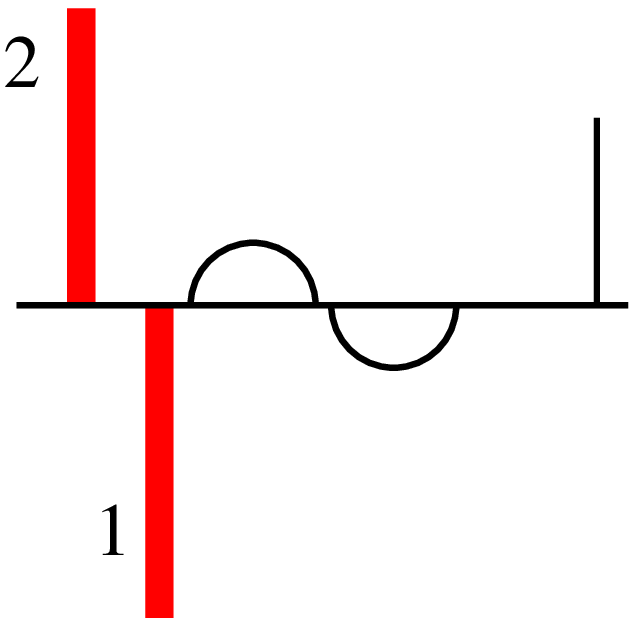}
      \end{varwidth}
      \,,  \\
      \label{U_5_2_3}
      U_{l=2}^{b=3} \left(
        \begin{varwidth}{\textwidth}
          \includegraphics[scale=0.35]{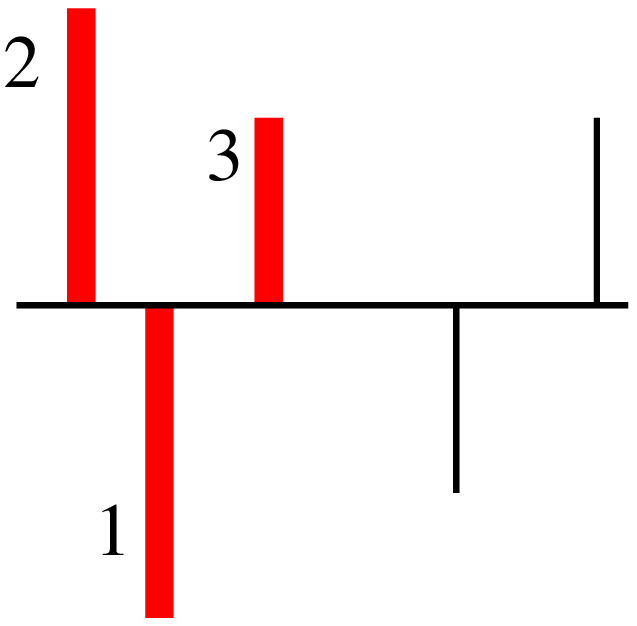}
        \end{varwidth}
      \right) & = &
      \begin{varwidth}{\textwidth}
        \includegraphics[scale=0.35]{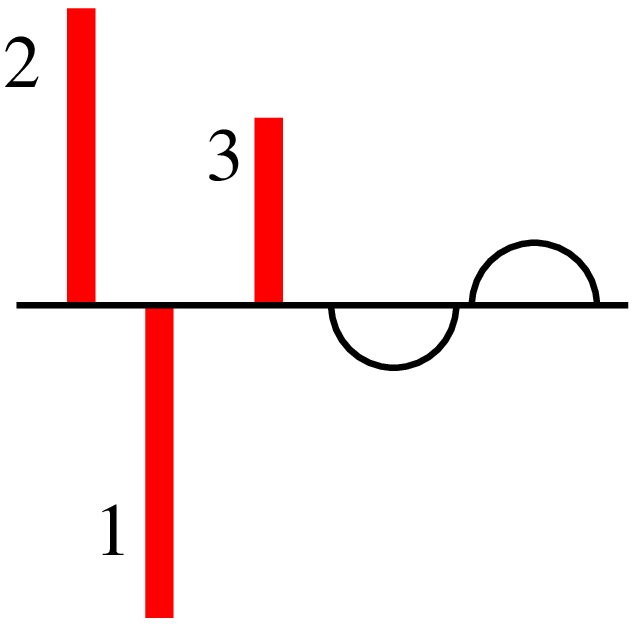}
      \end{varwidth}
      \,,  \\
      \label{U_5_2_2_2}
      U_{l=2}^{b=2} \left(
        \begin{varwidth}{\textwidth}
          \includegraphics[scale=0.35]{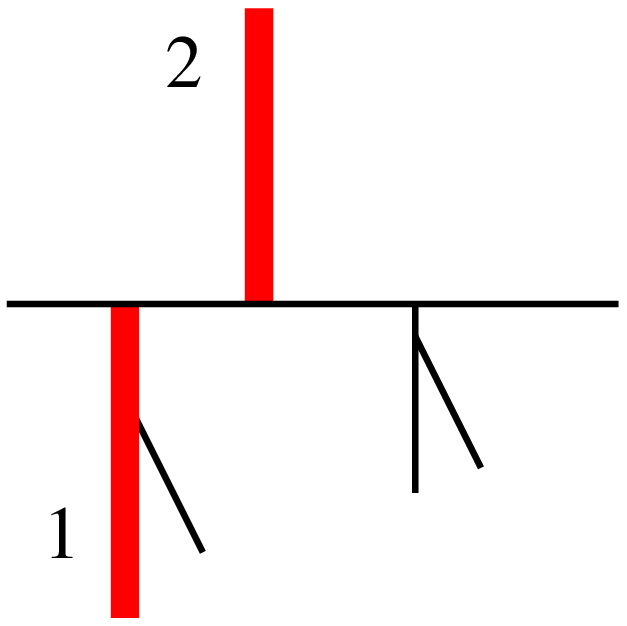}
        \end{varwidth}
      \right) & = &
      \begin{varwidth}{\textwidth}
        \includegraphics[scale=0.35]{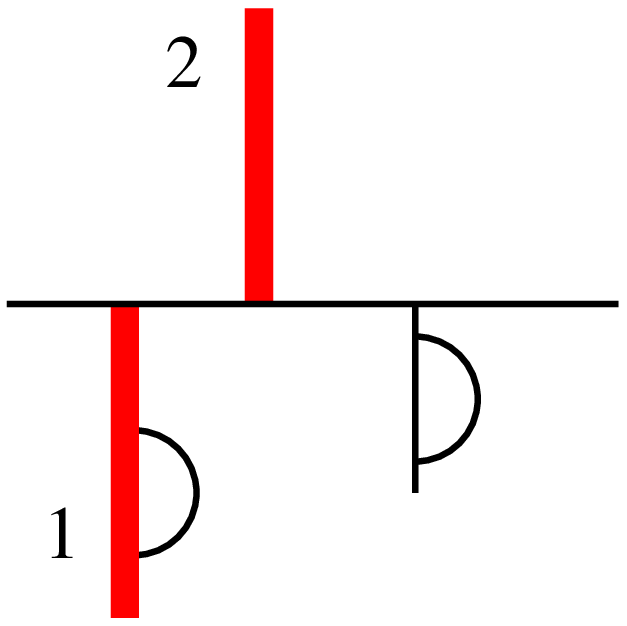}
      \end{varwidth}
      \,, \\
      \label{U_5_3_2}
      U_{l=3}^{b=2} \left(
        \begin{varwidth}{\textwidth}
          \includegraphics[scale=0.35]{figs/ex_b205.eps}
        \end{varwidth}
      \right) & = & 
      \begin{varwidth}{\textwidth}
        0
      \end{varwidth}
      \,.
    \end{eqnarray}
  \end{subequations}
}%
Eq.~(\ref{U_4_1_2}) gives an example of singly-looped configurations
(``1-loop diagrams'') generated by 
LoopSim when studying the $2\to4$ contributions to QCD dijet
production. Eq.~(\ref{U_4_2_2}) shows the ``2-loop diagrams'' generated
from the same event. The next equation shows what happens if we
add one
more particle to the final state. If, eq.~(\ref{U_5_2_3}), we now set
the number of Born 
particles for the same event to be $3$, we obtain only one $2$-loop
diagram instead of three, as represented in
eq.~(\ref{U_5_2_3}).\footnote{One might reasonably be surprised by
  this: after all, the result for the exact two-loop diagrams is
  independent of any choice of number of Born particles. The point is
  that if one studies soft and collinear corrections to the 3-jet
  cross section, then for the events in eq.~(\ref{U_5_2_2_1}) where the
  particle labelled $3$ in eq.~(\ref{U_5_2_3}) is virtual, the event
  will resemble a two-jet event and so not pass the 3-jet cuts.
  However if one studies the 3-jet cross section in a kinematic region
  where the cuts allow one of the jets to be much softer than the
  others, then to obtain sensible results it becomes necessary to use
  $b=2$ and include all diagrams on the right-hand side of
  eq.~(\ref{U_5_2_2_1}).  }
Finally, the last two examples of eq.~(\ref{eq:U-b-l-series}) give a
case with a splitting: the emitter is not looped, even if it is not a
Born particle.

We also give a few examples of the action of the $U_{\forall}^b$
operator:
{\allowdisplaybreaks
  \begin{subequations}
    \begin{eqnarray}
      U_{\forall}^{2} \left(
        \begin{varwidth}{\textwidth}
          \includegraphics[scale=0.35]{figs/ex_b201.eps}
        \end{varwidth}
      \right) & = &
      \begin{varwidth}{\textwidth}
        \includegraphics[scale=0.35]{figs/ex_b201.eps}
      \end{varwidth}
      -\; 
      \begin{varwidth}{\textwidth}
        \includegraphics[scale=0.35]{figs/ex_b202.eps}
      \end{varwidth}
      \;- \;
      \begin{varwidth}{\textwidth}
        \includegraphics[scale=0.35]{figs/ex_b203.eps}
      \end{varwidth}
      \;+ \;
      \begin{varwidth}{\textwidth}
        \includegraphics[scale=0.35]{figs/ex_b204.eps}
      \end{varwidth}
      \,,\\
      U_{\forall}^{2} \left(
        \begin{varwidth}{\textwidth}
          \includegraphics[scale=0.35]{figs/ex_b205.eps}
        \end{varwidth}
      \right) & = &
      \begin{varwidth}{\textwidth}
        \includegraphics[scale=0.35]{figs/ex_b205.eps}
      \end{varwidth}
      -\; 
      \begin{varwidth}{\textwidth}
        \includegraphics[scale=0.35]{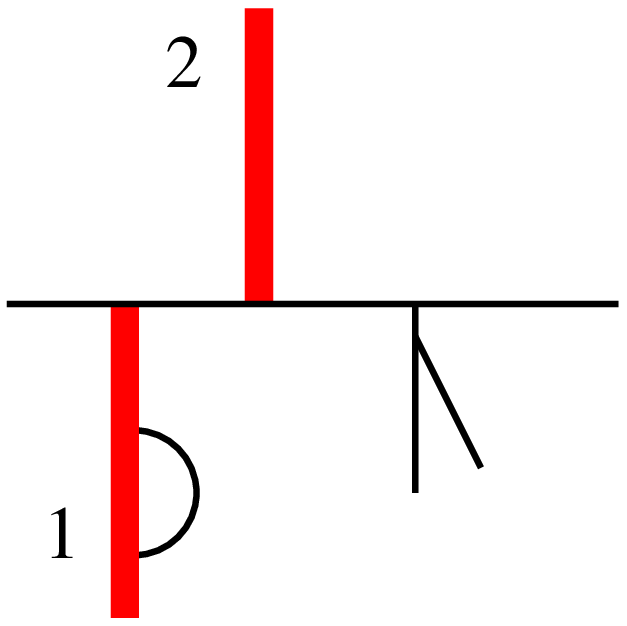}
      \end{varwidth}
      \;- \;
      \begin{varwidth}{\textwidth}
        \includegraphics[scale=0.35]{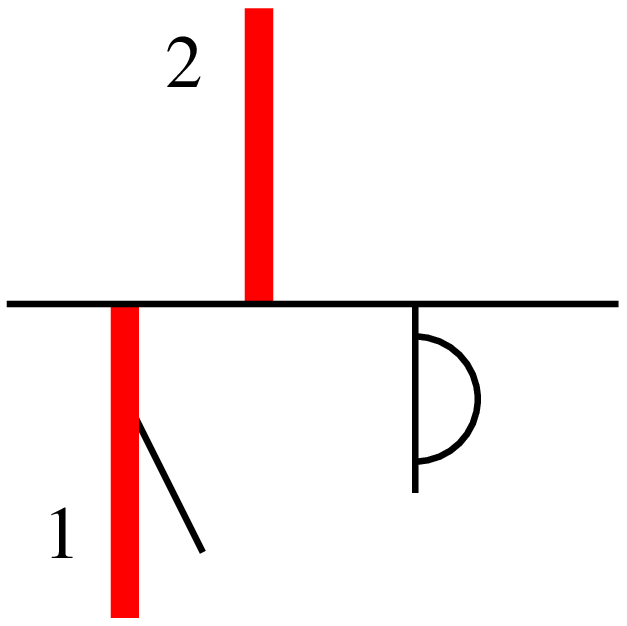}
      \end{varwidth}
      \;+ \;
      \begin{varwidth}{\textwidth}
        \includegraphics[scale=0.35]{figs/ex_b208.eps}
      \end{varwidth}
      \,,\;\; \\
      U_{\forall}^{2} \left(
        \begin{varwidth}{\textwidth}
          \includegraphics[scale=0.35]{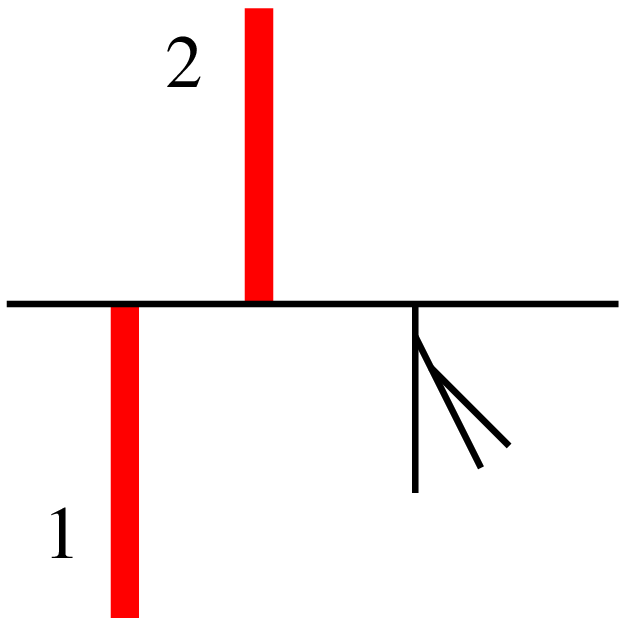}
        \end{varwidth}
      \right) & = &
      \begin{varwidth}{\textwidth}
        \includegraphics[scale=0.35]{figs/ex_b2secem01.eps}
      \end{varwidth}
      -\; 
      \begin{varwidth}{\textwidth}
        \includegraphics[scale=0.35]{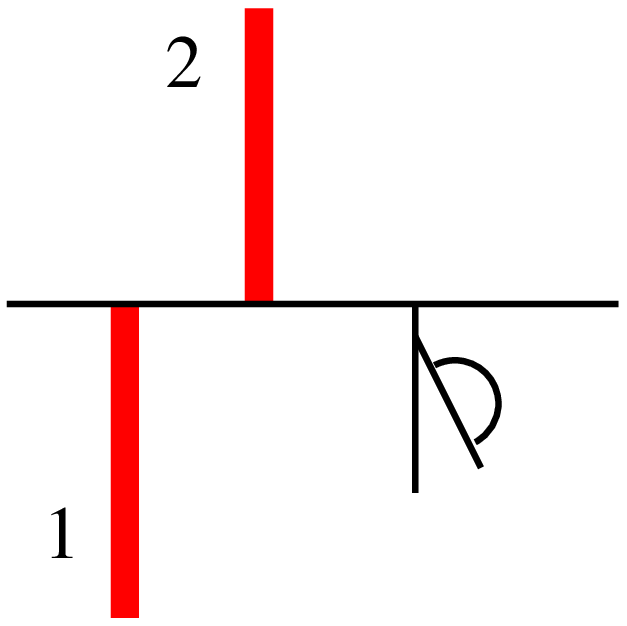}
      \end{varwidth}
      \,.
    \end{eqnarray}
  \end{subequations}
}%
In the last case, only one particle can become virtual because there
are two secondary emitters which cannot be looped.

\subsection{Treatment of flavour within LoopSim}
\label{sec:loopsim-flavour}

Let us now examine some of the issues that arise if we are to extend
the LoopSim method to processes with quarks and vector bosons.

\begin{figure}[t]
  \centering
  \includegraphics[scale=0.3]{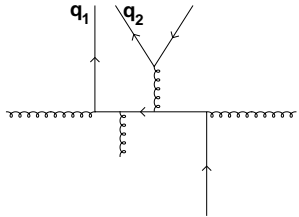}
  \caption{Example of an event where two quarks $q_1$ and $q_2$ may
    get recombined by the C/A algorithm.}
  \label{fig:quark_quark_recombination}
\end{figure}

We start with quarks and consider the situation depicted in
fig.~\ref{fig:quark_quark_recombination}. 
In this case, applying the C/A algorithm as in the previous section
will lead to the recombination of the two quarks $q_1$ and $q_2$,
which is clearly not physical.
If flavour information is available for events, then one can veto on
such a clustering, for instance by defining the clustering distance
$d_{qq}$ between two quarks to be infinite.
As discussed in \cite{Banfi:2006hf} such a modification alone is not
sufficient to systematically guarantee sensible treatment of flavour
in jet clustering.
Refs.~\cite{Banfi:2006hf,Hoeche:2009rj} have both discussed the
further modifications needed in the case of the $k_t$ algorithm.
A proper handling of flavour within LoopSim might seek to extend
those modifications to the C/A algorithm.
However, neither of the NLO programs that we use, MCFM and \nlojet,
provide information on particle flavours, so we defer such
modifications to future work and
just maintain the $d_{ij} = \Delta R_{ij}^2/R_{\LS}^2$ distance for
all partons.
For observables that are not flavour-sensitive this should not be a
major drawback, given the observation \cite{Banfi:2006hf} that
divergences associated with the mistreatment of flavour are strongly
subleading.
Were we to be interested in heavy tagged quarks, more careful
treatment might well be needed.
Note that there are also subtleties related to flavour and PDFs,
discussed in appendix~\ref{sec:incoming-partons}.

What about non-QCD particles, specifically vector bosons? 
Let us examine the case of Z~bosons.
A Z can be emitted from quarks or antiquarks and we would like this
to be reflected when establishing the approximate emission sequence,
because if the Z has been emitted from a quark, then that quark is a
secondary emitter and should not be looped.
In other cases a Z boson may be the hardest isolated object in an
event. Then it is to be considered a Born particle.
On the other hand we won't necessarily wish to consider diagrams where
a Z~boson is looped, because they would represent electroweak
corrections, not QCD corrections.

One issue in dealing with electroweak particles is that they are not
emitted from gluons. 
If one could distinguish between quarks and gluons, then this could be
accounted for during the C/A clustering, by defining the distance
$d_{gZ}$ between a Z and a gluon to be infinite.
Since we will not
know which partons are quarks or gluons, we adapt Frixione's isolation
procedure \cite{Frixione:1998jh} to decide if a Z boson relatively
close in angle to a parton $i$ is likely to have been emitted from
$i$. More precisely, if
\begin{equation}
  p_{ti} > \sqrt{p_{t\Z}^2+m_\Z^2}\frac{\Delta R_{iZ}}{R_{\LS}}\,,
\end{equation}
then we define $d_{iZ} = \Delta R_{iZ}^2/R_{\LS}^2$, otherwise
$d_{iZ}=\infty$. 
When recombining $i$ and Z into a particle $k$, then the identity
index $I_k$ is set equal to $I_i$ (a quark and a Z give a quark).
Our procedure means that a Z that is very collinear to a parton is always
considered to be emitted from that parton --- this makes sense because
such configurations are much more likely to occur when the parton is a
quark.
In contrast a soft parton in the general vicinity of a Z is not
clustered with the Z, which is sensible given that most soft partons
tend to be gluons.
Finally, for a recombination between a parton $i$ and
Z, we define the hardness of the branching $h_{iZ}$ as\footnote{This
  will be true in a future version of the code, but currently $h_{iZ}
  = \min(p_{ti}^2,p_{t\Z}^2)\Delta R_{iZ}^2$.}
\begin{equation}
  h_{iZ} \equiv \min(p_{ti}^2,p_{t\Z}^2+m_\Z^2)\frac{\Delta R_{iZ}^2}{R_{\LS}^2}\,,
\end{equation}
while a recombination of a Z with the beam has a hardness
\begin{equation}
  h_{ZB} \equiv p_{t\Z}^2+m_\Z^2\,.
\end{equation}
The latter means that for an event with just a parton and a recoiling
Z boson, the parton's beam hardness will always be lower than the Z's,
implying that for $b=1$ it is the Z-boson that will be the single Born
particle, as should be the case, at least when the parton has $p_t \ll
m_\Z$, i.e.\ in the kinematic regime that dominates the total cross
section for the Z.

Once a structure has been assigned to an event with a Z boson,
the next question is that of the looping procedure. When looping
partons it remains identical to before, with just a small extension of
the recoil procedure in order to deal with decay products of the Z
boson (see appendix~\ref{app:recoil_procedure}).
In the situations where the Z is not a Born particle (it is never an
emitter), straightforwardly following the procedure of
section~\ref{sec:looping-particles}, one would deduce that one should
loop the Z as well:
\begin{equation}
  \label{eq:looping-Z}
  U_{l=1}^{b=2} \left(
    \begin{varwidth}{\textwidth}
      \includegraphics[scale=0.35]{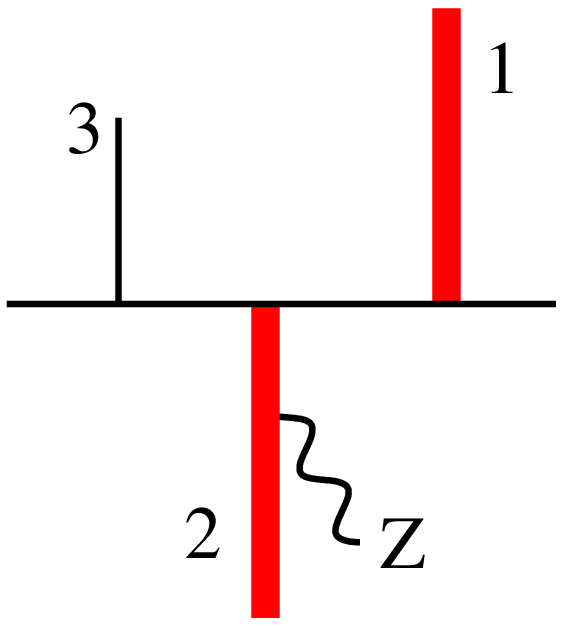}
    \end{varwidth}
  \right) = 
  -\; 
  \begin{varwidth}{\textwidth}
    \includegraphics[scale=0.35]{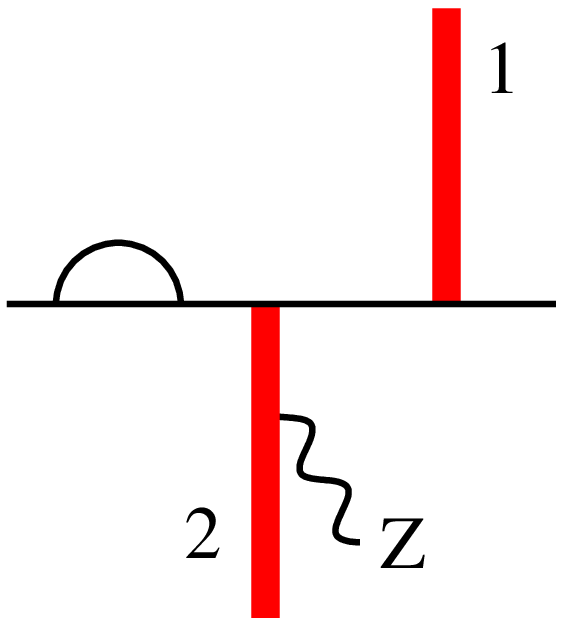}  
  \end{varwidth}
  \;- \;
  \begin{varwidth}{\textwidth}
    \includegraphics[scale=0.35]{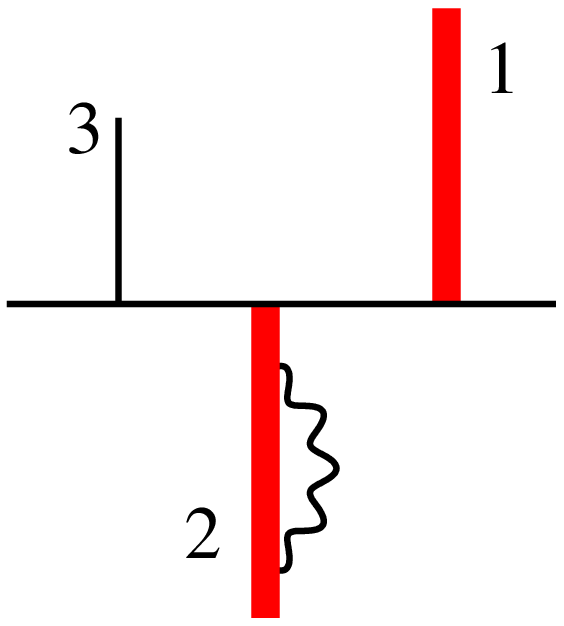}  
  \end{varwidth}
\end{equation}
(straight lines are partons, either quarks or gluons).
The rightmost diagram, with the looped Z, is not, however, a QCD
loop diagram: it is an electroweak loop correction to a multijet
event.
The LoopSim procedure does not aim to reproduce electroweak loop
corrections (though in this case it might be a reasonable
approximation). Furthermore, in any analysis that tags on Z bosons,
such a diagram would not be tagged and so would not contribute.
Thus, although the LoopSim procedure naturally generates events with
looped Z bosons, events like the rightmost diagram of
eq.~(\ref{eq:looping-Z}) are simply to be discarded.

For events with $\W^\pm$ bosons, the same procedure can be used as for
Z's. 
Note, however, that while the ``looped'' Z events may give a
reasonable approximation to actual electroweak loop diagrams, looped
$\W^\pm$ events will not. 
This is because $\W$-boson emission changes quark flavour: consider a
tree-level diagram $u\bar d \to b\bar b \W^+$, with the $\W^+$ emitted
collinearly off the incoming $\bar d$, converting it into a $\bar u$.
The LoopSim procedure would give a ``loop'' diagram $u\bar d \to b\bar
b$, with the $\W^+$ looped.
However no such loop diagram exists and the correct loop diagram would
instead involve $u\bar u \to b \bar b$.\footnote{A similar problem
  would appear to exist with the QCD diagram $u g \to b\bar b u$, with
  the incoming gluon splitting collinearly to give $u\bar u$, and the
  outgoing collinear $u$ being looped. Here, however, we are saved by
  the interplay between LoopSim and PDFs, as discussed in
  appendix~\ref{sec:incoming-partons}. It is crucial in this respect
  that all flavours that get looped are included in the PDFs.}
This is closely related to the phenomenon of Bloch-Nordsieck
violation~\cite{Ciafaloni:2000df} that is found when considering
electroweak double logarithms.
Since we in any case discard events in which electroweak bosons are
looped, this should not be a problem for the practical use of LoopSim
in events with $\W^\pm$ bosons.

\subsection{Merging NLO calculations and beyond}
\label{sec:loopsim-NLO}

Before explaining how we merge exact higher orders calculations, let
us mention how we use the LoopSim method in practice on tree-level
events at several different orders.
We introduce the notation X@\nbar$^p$LO to denote an approximation to
the N$^p$LO cross section for producing X, with all loop terms
estimated through the LoopSim procedure.
It is obtained by applying the $U_{\forall}^b$ operator to all
tree-level diagrams that can contribute up to N$^p$LO.
For instance, one can write
\begin{subequations}
  \label{eq:all-nLO-res}
  \begin{align}
    \mbox{Z@\nLO} & = U_{\forall}^1(\mbox{Z@LO})+U_{\forall}^1(\mbox{Z+j@LO})\,,\label{Z@nLO}\\
    \mbox{Z+j@\nLO} & = U_{\forall}^2(\mbox{Z+j@LO})+U_{\forall}^2(\mbox{Z+2j@LO})\,,\label{Z+j@nLO}\\
    \mbox{Z+j@\nnLO} & =
    U_{\forall}^2(\mbox{Z+j@LO})+U_{\forall}^2(\mbox{Z+2j@LO})+U_{\forall}^2(\mbox{Z+3j@LO})\,.\label{Z+j@nnLO}
  \end{align}
\end{subequations}
Notice that $U_{\forall}^1(\mbox{Z@LO})=\mbox{Z@LO}$ and
$U_{\forall}^2(\mbox{Z+j@LO})=\mbox{Z+j@LO}$. The terms
$U_{\forall}^1(\mbox{Z+j@LO})$ and $U_{\forall}^2(\mbox{Z+2j@LO})$
simulate up to one-loop corrections, and
$U_{\forall}^2(\mbox{Z+3j@LO})$ simulates up to two-loop corrections.

Now let us see how things work beyond tree-level
accuracy. We define $E_{n,l}$ to be a generic event at $l$ loops
(exactly calculated) with $n$ particles in the final state.
We first consider the case where only one-loop corrections are
computed exactly, so that we have tree-level events $E_{n,0}$ and
exact one-loop events $E_{n-1,1}$.
As before we can apply the unitarisation operator to the tree-level
events, $U_{\forall}^b(E_{n,0})$.
However, since we now include exact 1-loop contributions, $E_{n-1,1}$,
we must remove the approximate 1-loop contributions $U_{1}^b(E_{n,0})$
that are contained in $U_{\forall}^b(E_{n,0})$.
This alone is not sufficient, because among the extra
contributions from the exact 1-loop terms, there will be pieces that
are finite for a given $(n-1)$-parton configuration, but that can lead
to divergences when integrated over the $(n-1)$-parton phase space.
To cancel these extra divergences, we should introduce additional
approximate higher-loop contributions, which can be obtained by
applying the unitarisation operator $U_{\forall}^b$ to the difference
between the exact and approximate one-loop terms. So, rather than
including just events $E_{n-1,1}$ and subtracting $U_{1}^b(E_{n,0})$,
we include events $U_{\forall}^b(E_{n-1,1})$ and subtract
$U_{\forall}^b\left(U_1^b(E_{n,0})\right)$.
It is convenient to express this through a new operator
$U_{\forall,1}^{b}$ such that
\begin{subequations}
  \begin{align}
    U_{\forall,1}^{b}(E_{n,0}) & = U_{\forall}^b(E_{n,0})-U_{\forall}^b\left(U_1^b(E_{n,0})\right)\,,\label{U_1_loop_accuracy_l0}\\
    U_{\forall,1}^b(E_{n-1,1}) & =
    U_{\forall}^b(E_{n-1,1})\,,\label{U_1_loop_accuracy_l1}
  \end{align}
\end{subequations}
where the extra subscript $1$ on the $U_{\forall,1}^{b}$ indicates
that it is the form to use when the exact 1-loop result is to be
included.
The action of $U_{\forall,1}^b$ depends on the number of loops already
included in the event on which it operates: we subtract the one-loop
contribution returned by LoopSim only in tree-level events.
With this notation, one can compute the higher order corrections to
eqs.~(\ref{eq:all-nLO-res}) to one-loop accuracy,
\begin{subequations}
  \label{eq:nNLO-formulations}
  \begin{align}
    \mbox{Z@\nNLO} & = \mbox{Z@NLO} + U_{\forall,1}^1(\mbox{Z+j@NLO}_\only)\,,\\
    \mbox{Z+j@\nNLO} & = \mbox{Z+j@NLO} + U_{\forall,1}^2(\mbox{Z+2j@{N}LO}_\only)\,,\\
    \mbox{Z+j@\nnNLO} & = \mbox{Z+j@NLO} +
    U_{\forall,1}^2(\mbox{Z+2j@{N}LO}_\only)+U_{\forall,1}^2(\mbox{Z+3j@{N}LO}_\only)\,,
  \end{align}
\end{subequations}
where the ``only'' subscript on Z+$n$j@{N}LO$_\only$ means that we
take the highest order that contributes, i.e.\ here $\as^{n+1} \aEW$,
since the LO, $\as^{n} \aEW$, piece of Z+$n$j@{N}LO, is already taken
into account in the Z+$(n-1)$j@NLO contribution.
This implies that one should use consistent renormalisation
and factorisation scale choices across all different orders of the
calculation.
Note that in eq.~(\ref{eq:nNLO-formulations}) we have introduced the notation
\nbar$^p$N$^q$LO to denote an approximation to the N$^{p+q}$LO result
in which the $p$ highest loop contributions have been approximated
with LoopSim.

The extension of the procedure beyond one-loop accuracy is simple. For
instance, at two-loop accuracy, one has to subtract the approximated
two-loop contribution
$U_2^b(E_{n,0})-U_1^b\left(U_1^b(E_{n,0})\right)$ in
eq.~(\ref{U_1_loop_accuracy_l0}), and the other approximated two-loop
contribution $U_1^b(E_{n,1})$ in eq.~(\ref{U_1_loop_accuracy_l1}),
giving
\begin{subequations}
  \begin{align}
    U_{\forall,2}^b(E_{n,0}) & = U_{\forall}^b(E_{n,0})-U_{\forall}^b\left(U_1^b(E_{n,0})\right)-U_{\forall}^b\left[U_2^b(E_{n,0})-U_1^b\left(U_1^b(E_{n,0})\right)\right]\,,\label{U_2_loop_accuracy_l0}\\
    U_{\forall,2}^b(E_{n-1,1}) & = U_{\forall}^b(E_{n-1,1})-U_{\forall}^b\left(U_1^b(E_{n-1,1})\right)\,,\\
    U_{\forall,2}^b(E_{n-2,2}) & = U_{\forall}^b(E_{n-2,2})\,.
  \end{align}
\end{subequations}
Therefore, once Z+j@NNLO is calculated, one may compute for instance
\begin{equation}
  \mbox{Z@\nNNLO} = \mbox{Z@NNLO}+U_{\forall,2}^1(\mbox{Z+j@NNLO}_\only)\,.
\end{equation}
To be complete, let us mention the generalisation of our procedure to
$m$-loop accuracy
\begin{equation}
   U_{\forall,m}^b(E_{n-l,l}) =
   U_{\forall}^b(E_{n-l,l})+\sum_{j=1}^{m-l}\,(-1)^j\hspace{-1.2cm}
   \sum_{\scriptsize\hspace{0.4cm}
     \begin{array}{c}l_1,...,l_j\ge 1
       \\
       \hspace{0.15cm}l_1+\mathellipsis +l_j\le m-l\end{array}}\hspace{-1cm}
   U_{\forall}^b\circ U_{l_1}^b\circ\mathellipsis\circ U_{l_j}^b(E_{n-l,l})\,.
\end{equation}

We noted at the end of section~\ref{sec:looping-particles} that the
plain LoopSim procedure does not generate the finite terms needed to
cancel residual scale dependence from lower orders.
With the introduction of the exact loop contributions, those finite
terms do now get included. Thus for a given number of exact plus
simulated loops, as we increase the number of exact loops, we should
expect to see reductions in scale uncertainties.

\subsection{Expected precision of the method}
\label{sec:loopsim-expected-accuracy}

Let us briefly explain why the LoopSim method is expected to work in
the presence of giant $K$-factors.
We consider an observable $A$ computed respectively at NLO and
\nLO. We define $K^{(A)}_\text{NLO}$ such that
\begin{equation}
  \sigma_\text{NLO}^{(A)}=K^{(A)}_\text{NLO}\sigma^{(A)}_\text{LO}\,,
\end{equation}
and we assume that $K^{(A)}_\text{NLO}\gg 1$. This huge $K$-factor may come
from  logarithmic enhancements in the real NLO diagram or the
appearance of new scattering channels in the perturbative expansion.
The computation of $\sigma^{(A)}_\text{\nLO}$ gives the exact real
part of the NLO calculation as well as the divergent terms of the
virtual correction. Therefore
\begin{equation}
  \sigma^{(A)}_\text{\nLO}-\sigma^{(A)}_\text{NLO} = 
  {\cal O}\left(\as\sigma^{(A)}_\text{LO}\right)\,, 
\end{equation}
where, in writing $\order{\smash{\as \sigma^{(A)}_{\LO}}}$, we mean
that the term missing in the \nLO calculation, the finite part of the
1-loop correction, is not especially enhanced.
This leads to
\begin{equation}
  \sigma^{(A)}_\text{\nLO} = \sigma^{(A)}_\text{NLO}\left(1+{\cal O}\left(\frac{\as}{K^{(A)}_\text{NLO}}\right)\right)\,.
\end{equation}
The relative difference between the approximate and exact NLO
calculations is thus suppressed by the inverse $K$-factor.

Next, consider \nNLO accuracy.
The difference between $\sigma^{(A)}_\text{\nNLO}$ and
$\sigma^{(A)}_\text{NNLO}$ comes from the parts of the two-loop
corrections that are finite and associated with the LO topology, so
that they should be free of the enhancements that led to the large NLO
$K$-factor.
This implies 
\begin{equation}
  \sigma^{(A)}_\text{\nNLO} - \sigma^{(A)}_\text{NNLO} = {\cal O}\left(\as^2\sigma^{(A)}_\text{LO}\right)\,.
\end{equation}
If we define $K^{(A)}_\text{NNLO}$ such that $\sigma^{(A)}_\text{NNLO}
= K^{(A)}_\text{NNLO}\sigma^{(A)}_\text{LO}$, then we can write
\begin{equation}
  \sigma^{(A)}_\text{\nNLO} = \sigma^{(A)}_\text{NNLO}\left(1+{\cal O}\left(\frac{\as^2}{K^{(A)}_\text{NNLO}}\right)\right)\,.
\end{equation}
If $K^{(A)}_\text{NLO}\gg 1$, one can expect $K^{(A)}_\text{NNLO}\gg 1$ too.

\section{The reference-observable method}
\label{sec:alternative_method}

Given the novelty of the LoopSim method, it is useful to have an
alternative way of estimating the size of the NNLO contributions that
we will approximate with LoopSim.
Here we outline such an alternative method, which, though less
flexible than the LoopSim approach, will provide a valuable
cross-check and help us build our confidence in results of the LoopSim
method.

Let us explain it for observables in the Z+j process. 
Our aim is to estimate $\sigma^{(A)}_\text{NNLO}$ for some observable
$A$.\footnote{More precisely, $\sigma^{(A)}$ is the cross section for the
  observable $A$ to pass some given cuts; it is only for brevity that
  we use here the somewhat inaccurate shorthand ``cross section for
  observable $A$''.}
We assume that we have a reference observable which is identical to
the observable $A$ at LO.
For instance, one might consider $\reference = p_{t,\Z}$ and
$A=p_{t,j}$.
We can write the NNLO Z+j prediction for $A$ in terms of the NNLO
prediction for the reference observable plus the NLO Z+2j difference
between $A$ and the reference cross section
\begin{subequations}
  \begin{align}
    \label{eq:ref-meth-exact}
    \sigma^{(A)}_\text{Z+j@NNLO} &=
    \sigma^{(\reference)}_\text{Z+j@NNLO} +
    (\sigma^{(A)} - \sigma^{(\reference)})_\text{Z+j@NNLO}\,,\\
    &= \sigma^{(\reference)}_\text{Z+j@NNLO} + (\sigma^{(A)} -
    \sigma^{(\reference)})_\text{Z+2j@NLO}\,.
  \end{align}
\end{subequations}
The second equality is possible because 2-loop NNLO corrections to Z+j
have the topology of Z+j at LO. Therefore, their contributions to
the observables $A$ and ref are identical and cancel in the difference in
eq.~(\ref{eq:ref-meth-exact}).
 
If we have reason to believe that the perturbative expansion for the
reference observable converges well, we can conclude that
$\sigma^{(\reference)}_\text{Z+j@NNLO}
-\sigma^{(\reference)}_\text{Z+j@NLO}$ is genuinely a small
correction.
Then
\begin{equation}
  \label{eq:ref-nNLO-final}
  \sigma^{(A)}_\text{Z+j@NNLO} \simeq \sigma^{(\reference)}_\text{Z+j@NLO} +
  (\sigma^{(A)} - \sigma^{(\reference)})_\text{Z+2j@NLO}\,,
\end{equation}
\ie we approximate the NNLO distribution for $A$ in terms of the NLO
distribution for the $\reference$ observable and a NLO calculation for
difference between the $A$ and $\reference$ distributions, both of
which are exactly calculable.
The missing part is suppressed by a relative factor $1/K^{(A)}$, as
for the LoopSim 
method. For Z+j, one can see from fig.~\ref{fig:LO_NLO_dist} that
$p_{t,\Z}$ seems to be an acceptable reference observable for $p_{t,j}$
and $H_{T,\jets}$.

In the sections that follow we shall, for brevity, refer to the RHS of
eq.~(\ref{eq:ref-nNLO-final}) as ``ref.\ \nNLO'' even though it does
not quite adhere to our the meaning of \nNLO as set out in section 2,
i.e.\ in terms of the specific sets of tree-level and loop diagrams
that are included exactly.

\section{Validation: comparison to DY at NNLO}
\label{sec:DY-comparison}

The cross section for the Drell-Yan process is known with exclusive
final states up to NNLO accuracy \cite{Melnikov:2006kv, Catani:2009sm}. 
Above a certain value of lepton transverse momentum, one finds giant
corrections to the lepton $p_t$ spectra when going from LO to NLO and
large ones from NLO to NNLO.  
This gives us an opportunity to directly test the performance of the
LoopSim method by comparing its \nNLO results to exact NNLO spectra
for lepton pair production.

Before examining \nNLO results, it is useful to compare \nLO with
NLO. 
If they are in reasonable agreement for some observable, then that
serves as a first indication that the LoopSim estimate of missing loop
corrections is sensible for that observable.

\begin{figure}[t]
  \centering
  \scalebox{0.55}{\includegraphics{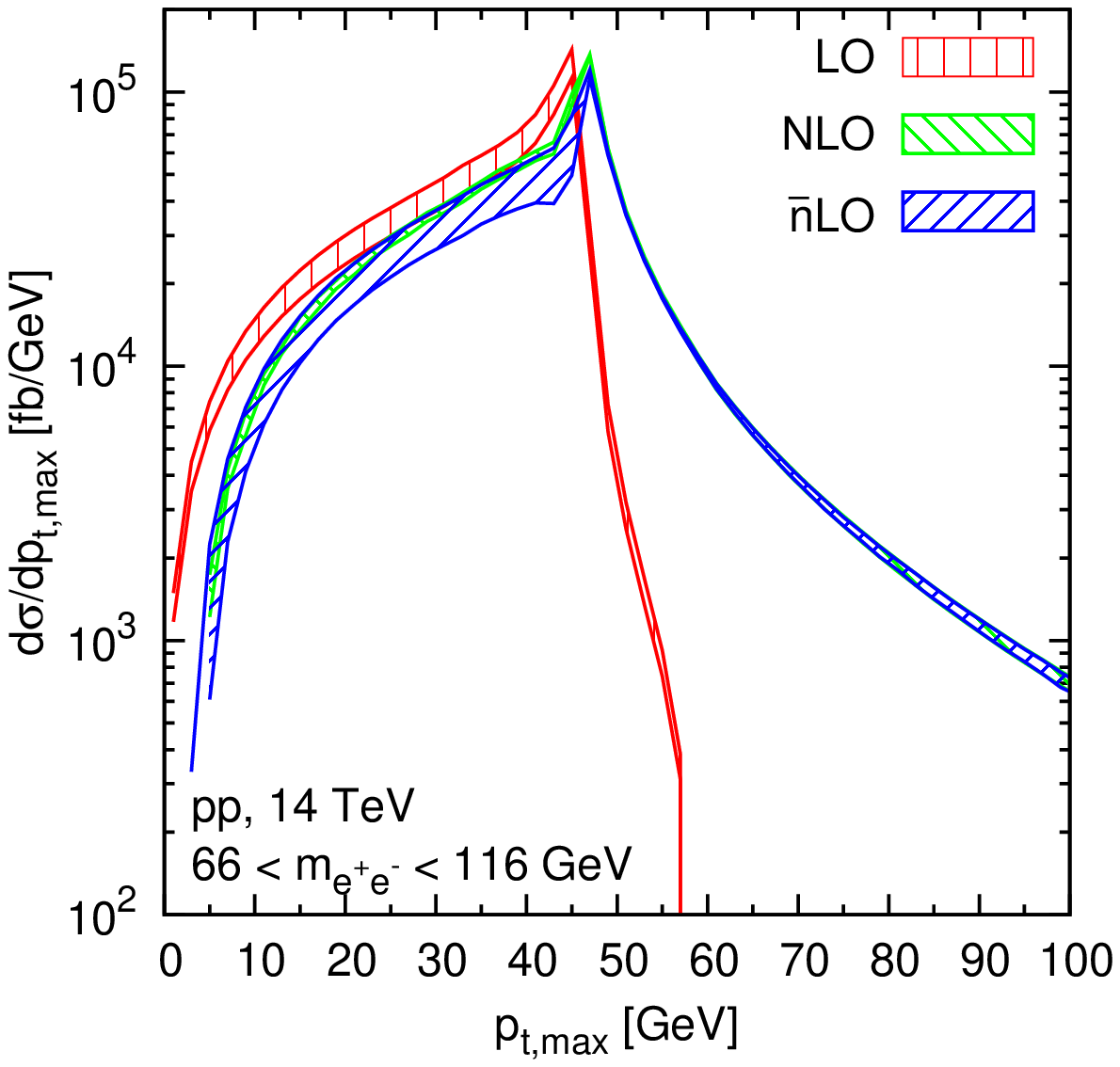}}\;\;
  \hspace{20pt}
  \scalebox{0.55}{\includegraphics{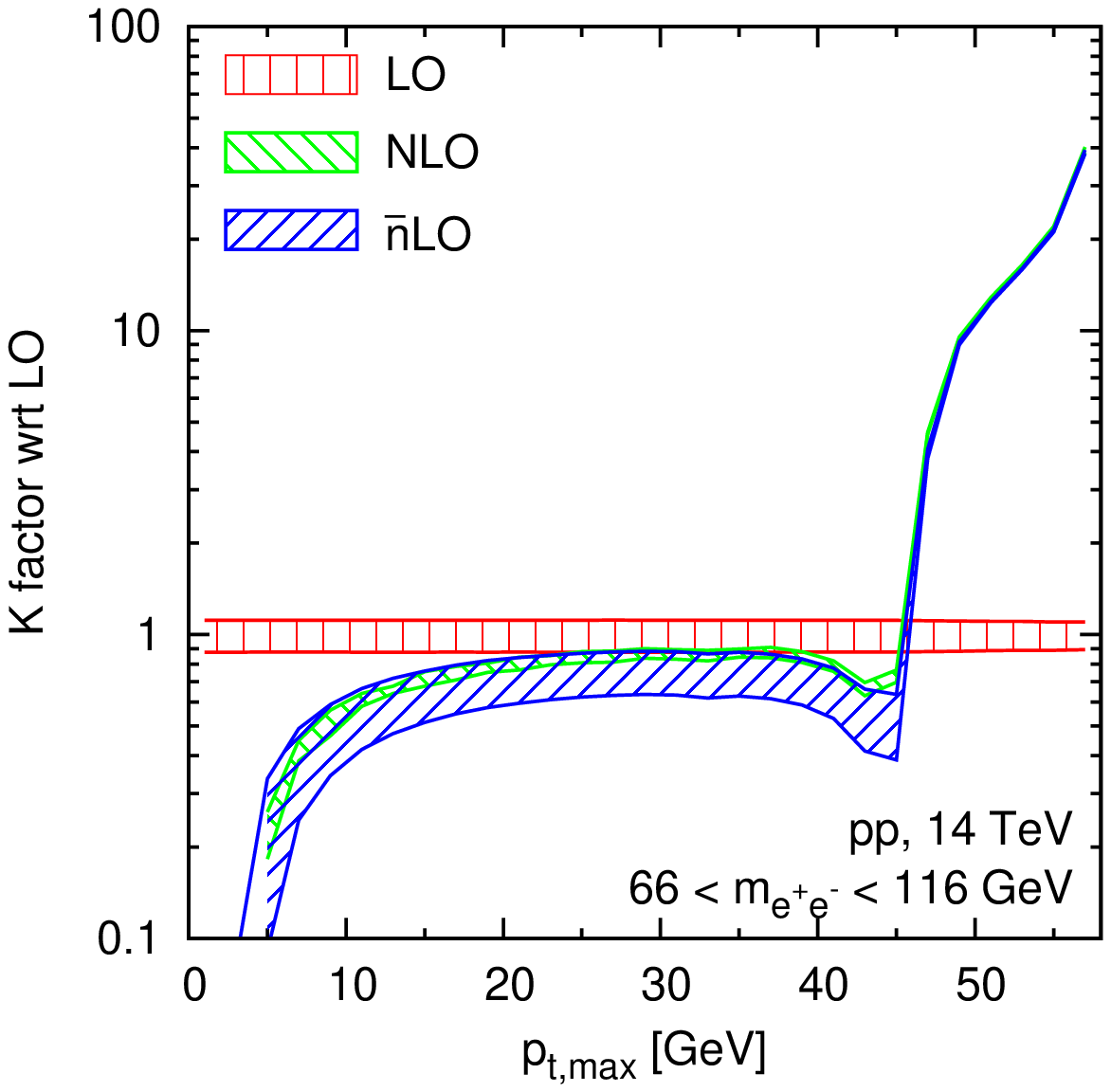}}\;\;
  \caption{%
  Comparison between \nLO results from LoopSim (with $R_{\LS}=1$) and
  exact NLO results for the Drell-Yan process. 
  The left-hand plot shows the  transverse momentum spectrum of the
  harder lepton, while the right-hand plot gives the corresponding K
  factors w.r.t. LO. 
  The uncertainty bands were obtained by varying $\muR=\muF$ by a factor of
  $\frac12$ and $2$ around a default choice of $m_\Z$.
}
  \label{fig:dy-nlo}
\end{figure}

Fig.~\ref{fig:dy-nlo} gives the comparison of the \nLO, NLO and LO
results for the production of an $e^+e^-$ pair within the mass window
of $66 < m_{e^+e^-} < 116$~GeV at a proton-proton centre of mass
energy of~$14$~TeV.
The left-hand plot shows the cross section differential in the
transverse momentum of the harder of the two leptons.
The right-hand plot gives the corresponding K factor with respect to
LO.
The results were obtained with MCFM~5.3~\cite{Campbell:2000bg,mcfm},
with its default set of electroweak parameters and NNLO MSTW2008
parton distribution functions.
The uncertainty bands in Fig.~\ref{fig:dy-nlo} correspond to varying the
renormalisation and factorisation scales $\muR=\muF$ by a factor of
$\frac12$ and $2$ around a default choice of $m_\Z$.
In the \nLO result we fixed the value of the LoopSim radius parameter
to be $R_{\LS}=1$, which naturally places interparticle and
particle-beam clustering on the same footing (though the \nLO result
here is actually independent of $R_{\LS}$, because there is at most one
isolated QCD parton in the final state).

There are three relevant regions of transverse momentum in
fig.~\ref{fig:dy-nlo}.
For $p_{t,\max} \lesssim \frac12 m_\Z$ (low $p_t$) the distribution is
dominated by on-shell Z-bosons and its shape is governed by the
angular distribution of the Z decays in their centre-of-mass frame.
The peak close to $\frac12 m_\Z$ corresponds to Z-bosons that decay in
a plane at right-angles to the beam.
For $\frac12 m_\Z \lesssim p_{t,\max} < 58\GeV$ (intermediate $p_t$), the LO
distribution comes from Z-bosons that are off shell, which allows
the $p_t$ of the lepton to be larger than $\frac12 m_\Z$.
The narrow width of the Z causes the distribution to fall very
steeply.
The $58\GeV$ upper edge of this region is a consequence of our cut
on $m_{e^+e^-} < 116\GeV$.
Above $58\GeV$ (high $p_t$) the LO distribution is zero.

In the low $p_t$ region, the NLO correction is moderate and
negative. There is no strong reason to believe that the LoopSim method
should work here, but it turns out that the \nLO result reproduces the
structure of the correction, even if its scale dependence remains much
larger than that of the NLO result (this is because the LoopSim
procedure does not include the finite terms that would partially
cancel the LO scale dependence).
In the intermediate $p_t$ region, we see a ``giant'' NLO $K$-factor. 
It comes about because initial-state radiation can give a boost to the
Z-boson, causing one of the leptons to shift to higher $p_t$ (it
becomes the ``max'' lepton).
The spectrum of QCD radiation falls much less steeply than the
Z-boson lineshape, so this NLO correction dominates over the LO
result.
In this region the exact loop correction, proportional to the LO
result, becomes almost irrelevant and we see near perfect agreement
between \nLO and NLO.
In the high-$p_t$ region only the real emission diagrams of Z@NLO
contribute and  \nLO becomes identical to NLO (both correspond to the
Z+j@LO result).
Similar results hold for the $p_{t,e^\pm}$ distribution, while the
$p_{t,\text{min}}$ lacks the giant $K$-factor in the intermediate
region.

\begin{figure}[t]
  \centering
  \scalebox{0.55}{\includegraphics{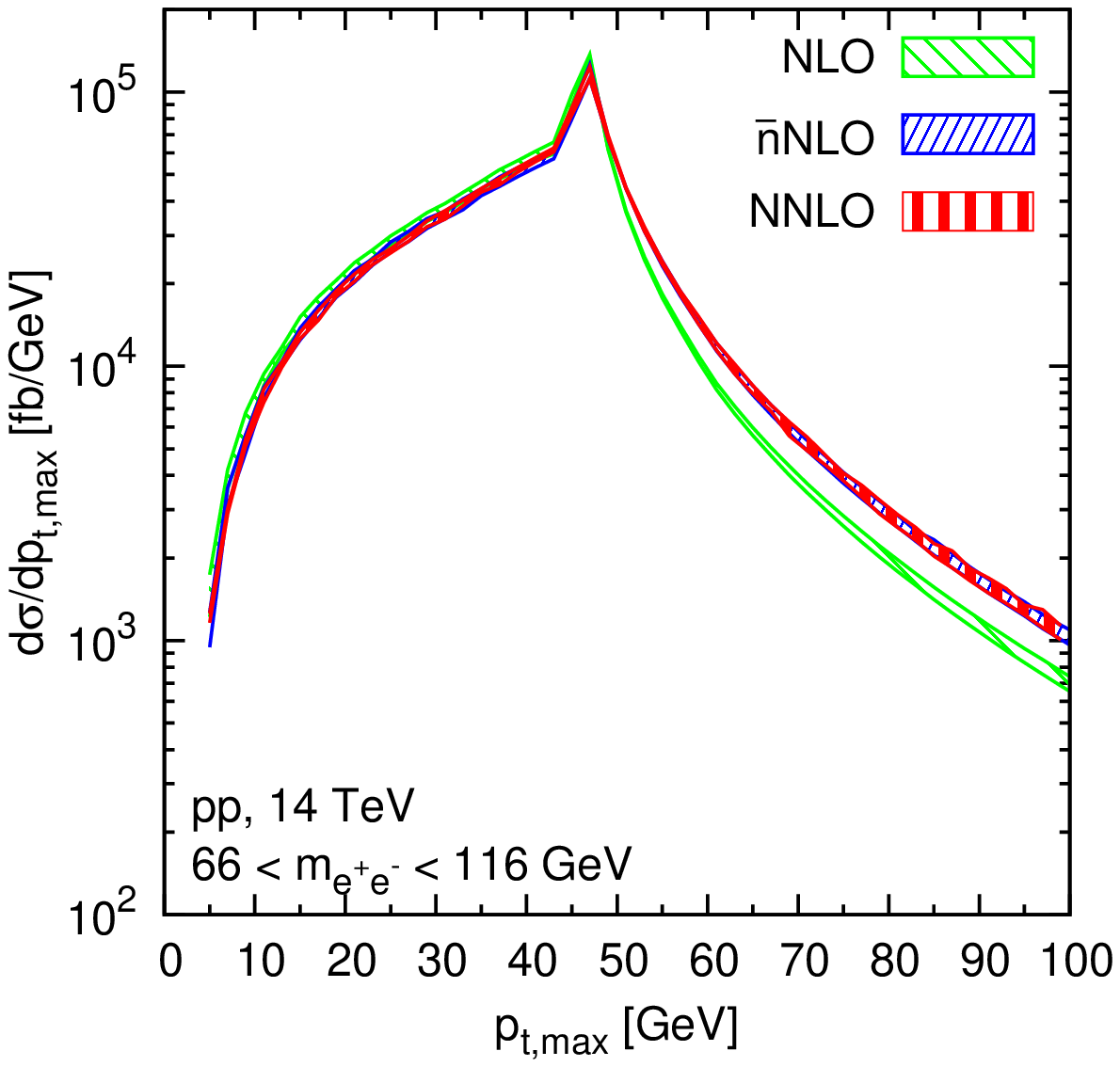}}\;\;
  \hspace{20pt}
  \scalebox{0.55}{\includegraphics{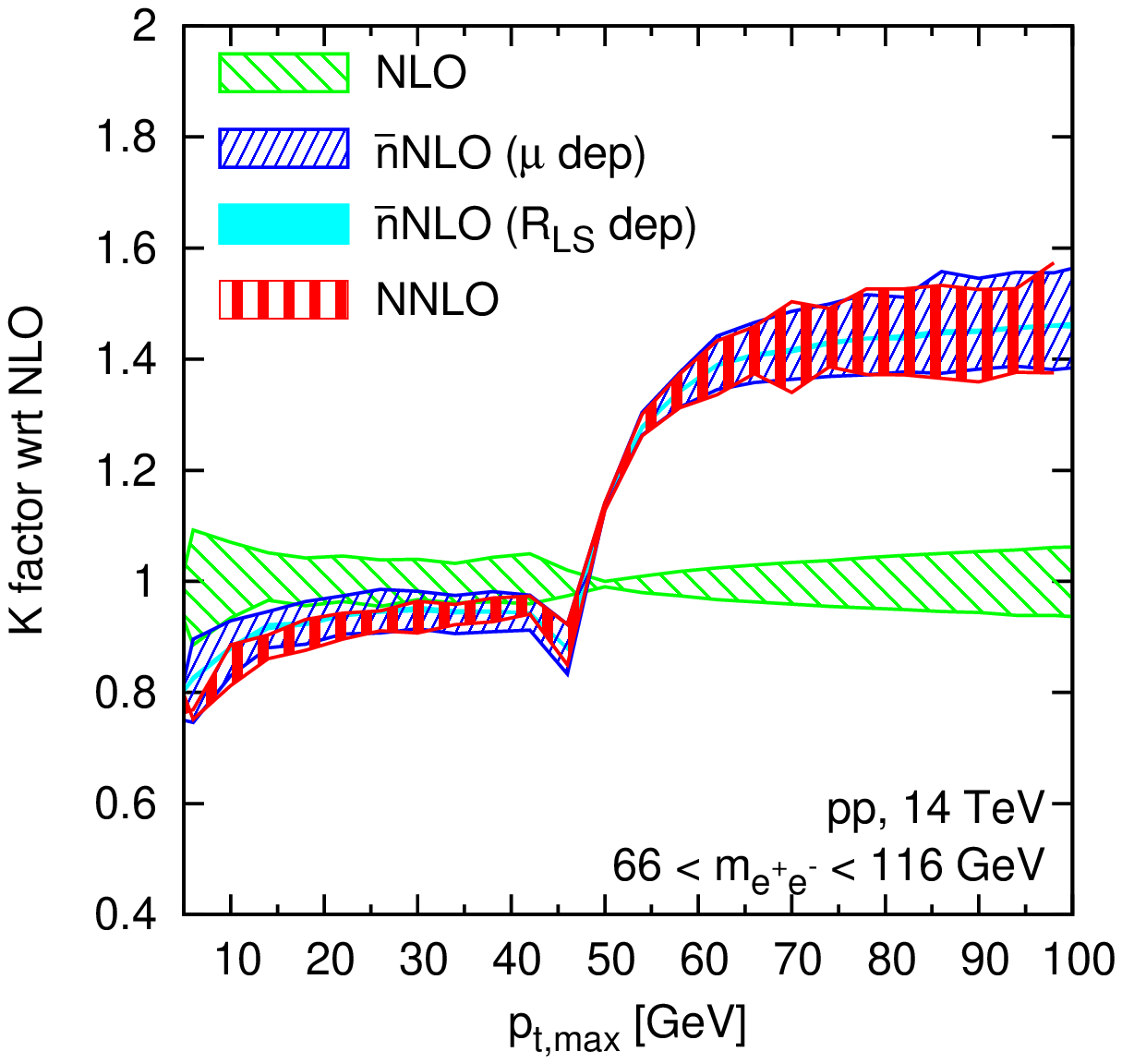}}\;\;
  \caption{%
  Comparison between \nNLO results from LoopSim+MCFM (with $R_{\LS}=1$) 
  and full NNLO results for the Drell-Yan process from DYNNLO.  
  The left-hand plot shows the transverse momentum spectrum of the
  harder lepton,  
  while the right-hand plot gives the corresponding K factors
  w.r.t. NLO.   
  The uncertainty bands come from varying the factorisation and renormalisation
  scales by factors 1/2~and~2.  
  In the right-hand plot we also show the (thin) band related to
  changing the \nNLO $R_{\LS}$ parameter from 0.5 to 1.5, at fixed
  $\muR=\muF=m_\Z$.  }
  \label{fig:dy-nnlo}
\end{figure}

A similar comparison between \nNLO and NNLO spectra is shown in
fig.~\ref{fig:dy-nnlo}. The NNLO results were obtained with DYNNLO 1.0
\cite{Catani:2009sm,Catani:2007vq,dynnlo}, used with a set of electroweak
parameters compatible with that of MCFM.\footnote{
  In its $\order{\as}$ and $\order{\as^2}$ contributions, DYNNLO
  includes among its parameters a cut on the $p_t$ of the Z boson.
  The cut is applied to both real and virtual terms and its impact
  should vanish as it is taken towards zero.
  It is, however, required to be non-zero for the numerical stability of the
  MCFM Z+j NLO calculation that is among the components of DYNNLO.
  We set the cut equal to $0.1\GeV$ in the $\order{\as}$ term and to 
  $1\GeV$ in the $\order{\as^2}$ term.
  A related $1\GeV$ cut was placed on the $\order{\as^2}$ piece of
  the \nNLO result (while none was used at $\order{\as}$).
  The impact of the $1\GeV$ cut is small but not entirely negligible
  close to the peak (where, physically, NNLO should in any case be
  supplemented with a resummation).
} 

In the low-$p_t$ region we find quite good agreement between the \nNLO
and NNLO results (with somewhat larger uncertainty bands for \nNLO).
Such a result was not guaranteed a priori, even if it is not entirely
surprising given the reasonable agreement that we saw between \nLO
and NLO.
In the intermediate $p_t$ region, where the NNLO/NLO corrections are
substantial, the agreement is excellent. This was expected.
At high $p_t$ the agreement should be exact, and does seem to be, 
within statistical fluctuations.
The dependence on $R_{\LS}$ (shown in the right-hand plot) has been
estimated by varying its value from 0.5 to 1.5. The effects are small.

Finally, we note that similar features and a similar level of
agreement between \nNLO and NNLO are to be found in the
$p_{t,\text{min}}$ and $p_{t,e^\pm}$ distributions.

\section{Results for the $\Z+$jet process}
\label{sec:results}

In the previous section, we studied the Z production process and showed that
our procedure correctly reproduces the $p_t$ distribution of the hardest
lepton at NNLO, even, unexpectedly, in regions where the $K$-factor is not large.
In this section we study the Z+j process, whose NNLO cross-section is not
known yet, but which leads to giant $K$-factors at NLO for some observables
as explained in the introduction.
Therefore, their NNLO contributions are expected to be accurately described by
the LoopSim method.
Throughout this section we use MCFM 5.7, including the Z+2j process at
NLO~\cite{Campbell:2003hd}, with the NLO CTEQ6M PDFs.
We will take three
different values for the renormalisation and factorisation scales:
$\muR = \muF = \frac12\mu_0$, $\mu_0$ and $2\mu_0$, with
\begin{equation}
  \mu_0 = \sqrt{m_\Z^2+p_{t,j1}^2}\,,
\end{equation}
where $p_{t,j1}$ is the transverse momentum of the hardest jet.
At high $p_t$, this scale choice should be quite similar to that used
in \cite{Berger:2010vm} and has the same $p_t$ scaling as those in
\cite{Bauer:2009km,Denner:2009gj}.
The $R_{\LS}$ uncertainty is measured at $\muR=\muF=\mu_0$ using three
different values for it: $R_{\LS} = 0.5$, $1$, $1.5$.

In addition to the 3 observables shown in the introduction, $p_{t,\Z}$,
$p_{t,j1}$ and $H_{T,\jets} = \sum_{i=1}^\infty p_{t,j i}$, we will
also consider
\begin{equation}
  H_{T,\tot}\equiv H_{T,\jets}+p_{t,\Z}\,.
\end{equation} 
We only include events for which $p_{t,j1} > 200\GeV$.

\subsection{Validation at \nLO}
\label{sec:nLO}

\begin{figure}
  \centering
  \scalebox{0.8}{\includegraphics{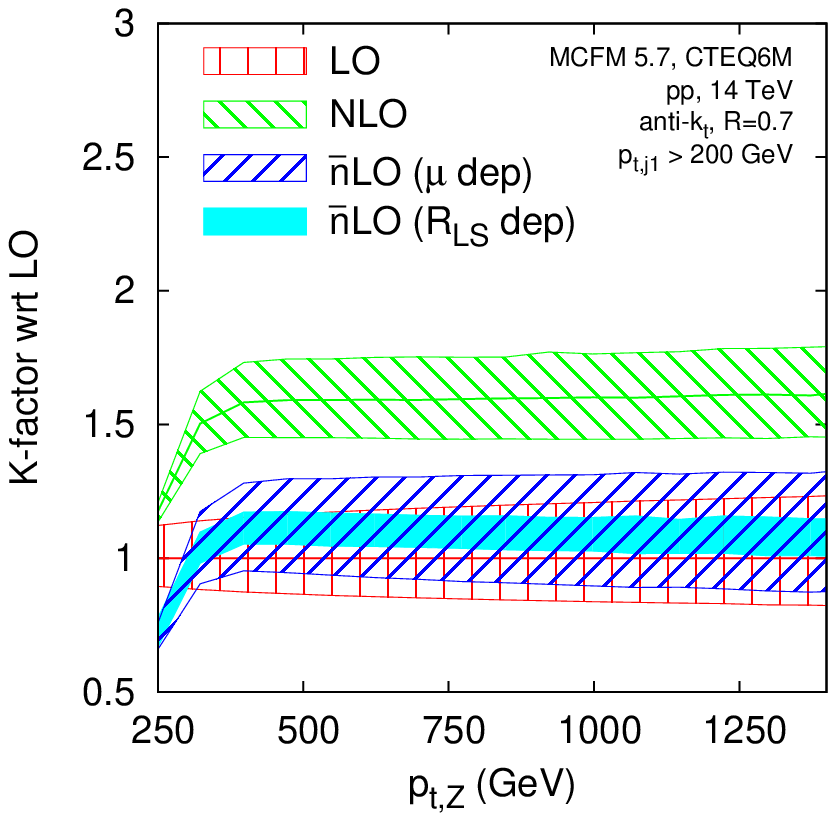}}\;\;
  \scalebox{0.8}{\includegraphics{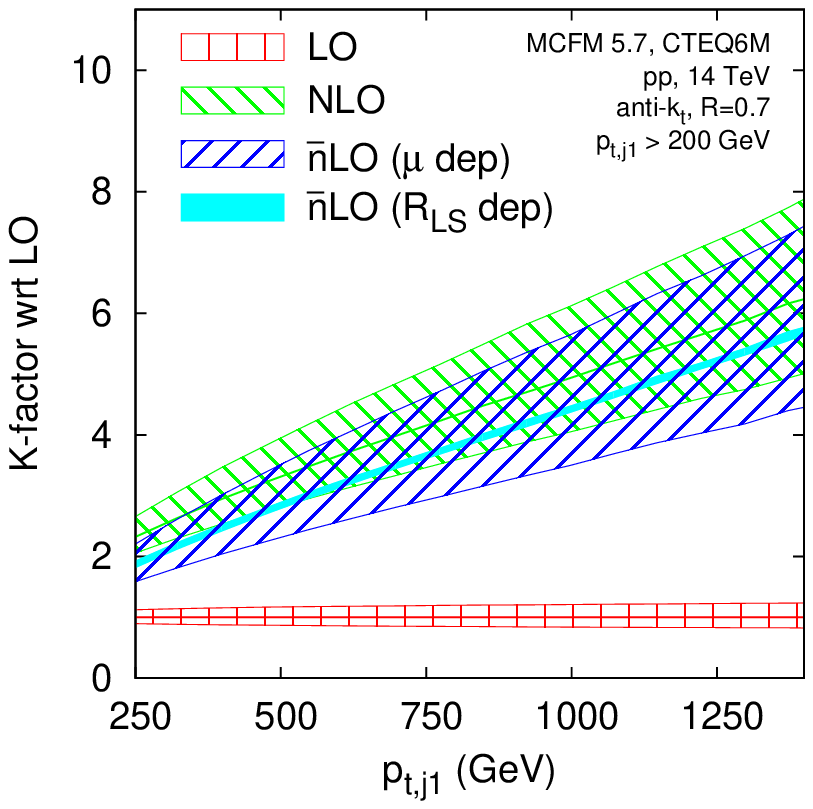}}\\
  \vspace{0.5cm}
  \scalebox{0.8}{\includegraphics{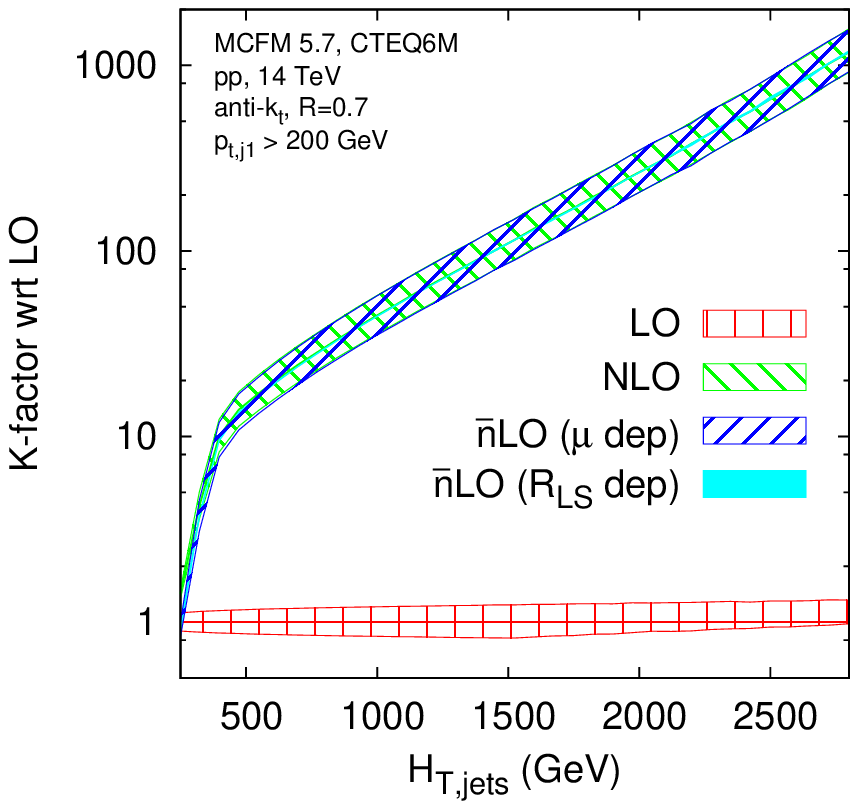}}\;\;
  \scalebox{0.8}{\includegraphics{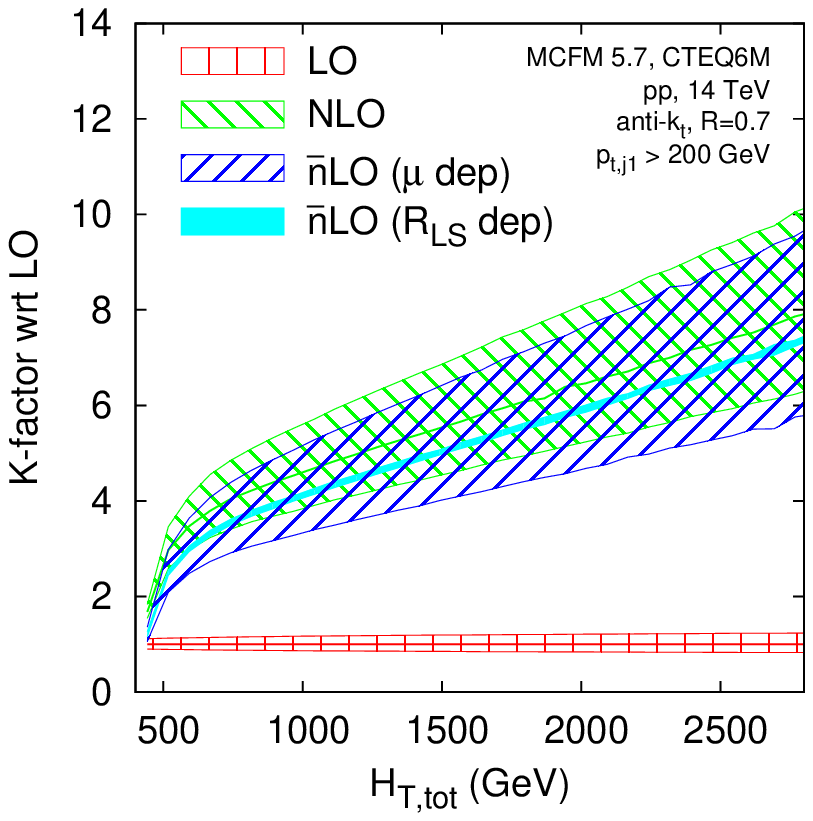}}
  \caption{Comparison of the \nLO/LO $K$-factor with the NLO/LO
    $K$-factor, together with their scale and $R_{\LS}$ uncertainties 
    for four observables in the Z+jet process.}
  \label{fig:nLO_NLO}
\end{figure}

As a first investigation of the performance of the LoopSim method, let
us examine how the \nLO approximation compares to the full NLO result.
Fig.~\ref{fig:nLO_NLO} shows the $K$-factors for the \nLO and NLO
predictions, with uncertainty bands from scale and $R_{\LS}$ variations.

In the upper-left plot, one sees that the \nLO prediction for the $p_{t,\Z}$ 
distribution gives a somewhat smaller $K$-factor than the NLO result. 
We interpret this as being because certain genuine loop effects are
not taken into account  
by the LoopSim method, for example those related to threshold logarithms, 
which depend crucially on the factorisation scheme of the parton 
distribution functions.
The \nLO result does, however, reproduce the $p_t$ dependence of the
$K$-factor, \ie the dip towards $p_t = 200\GeV$. This dip arises because
of the requirement in our event selection that there should be at least
one jet with $p_t > 200\GeV$.
At LO this induces a step-function in the $p_{t,\Z}$ distribution at
$200\GeV$.
At NLO, soft and collinear emissions smoothen out that threshold and
the \nLO calculation correctly reproduces the resulting interplay between 
real and virtual terms.

In the three remaining plots of fig.~\ref{fig:nLO_NLO}, for
$p_{t,j1}$, $H_{T,\jets}$ and $H_{T,\tot}$, all of which have giant
$K$-factors, one sees good agreement between the \nLO and NLO results.
This is because the dominant NLO contribution comes from events in the
B and C-type configurations of fig.~\ref{fig:z+jet-diags}, for which
there is no corresponding QCD 
loop correction. The LoopSim method merely serves to cancel the
divergences that arise from soft and collinear emissions off $A$-type
configurations and these are not dominant overall.

The $R_{\LS}$ dependence, also shown on these four plots, only comes 
from $1$-loop events generated by LoopSim.
Therefore, for an observable $A$ studied in Z+j@\nLO with two different
values $R_0$ and $R_1$ for $R_{\LS}$, one can write:
\begin{equation}
  \label{eq:RLS-uncert}
  \sigma^{(A)}_\text{Z+j@\nLO,$R_1$}-\sigma^{(A)}_\text{Z+j@\nLO,$R_0$}
  = 
  \sigma^{(p_{t,\Z})}_\text{Z+j@\nLO,$R_1$}-\sigma^{(p_{t,\Z})}_\text{Z+j@\nLO,$R_0$}
%
\end{equation}
as long as $A$ coincides with $p_{t,\Z}$ at LO (it does for each of 
$p_{t,j1}$, $H_{T,\jets}$ and $\frac12 H_{T,\tot}$).
This means that the {\it absolute} uncertainty due to $R_{\LS}$ is the
same for $A$ and $p_{t,\Z}$. Therefore, the {\it relative} uncertainty
due to $R_{\LS}$ is expected to be roughly inversely proportional to
the $K$-factor for $A$, in analogy with the discussion of
sec.~\ref{sec:loopsim-expected-accuracy}.
This explains why the $R_{\LS}$ dependence (solid cyan band) looks
significantly smaller for $p_{t,j1}$, $H_{T,\jets}$ and $H_{T,\tot}$
than it does for $p_{t,\Z}$ plot.
%

\subsection{Results at \nNLO}
\label{sec:nNLO}

\begin{figure}
  \centering
  \scalebox{0.8}{\includegraphics{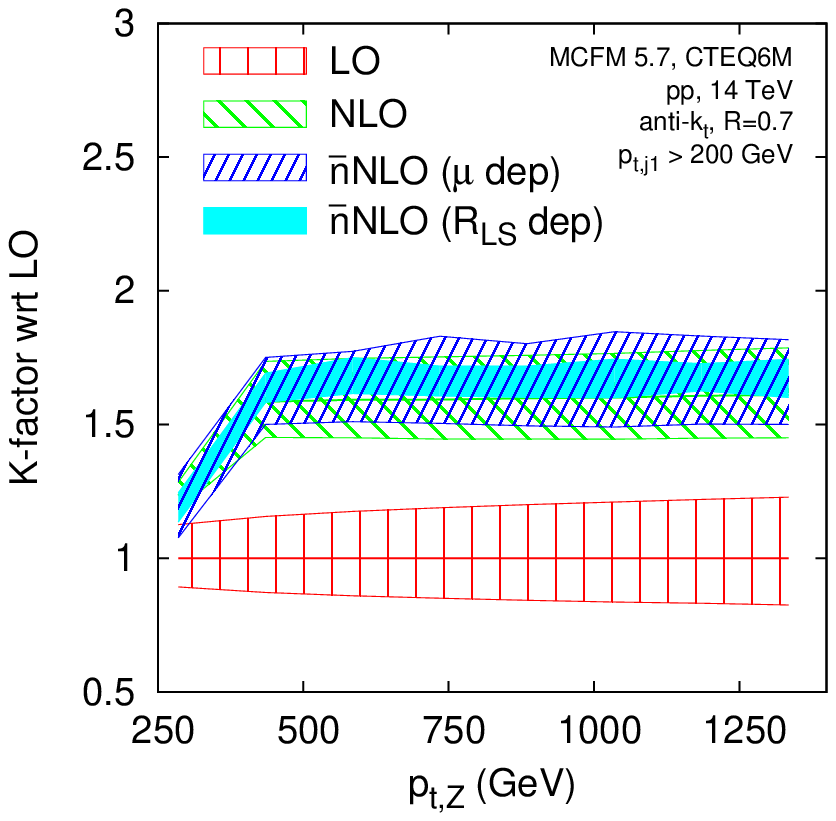}}\;\;
  \scalebox{0.8}{\includegraphics{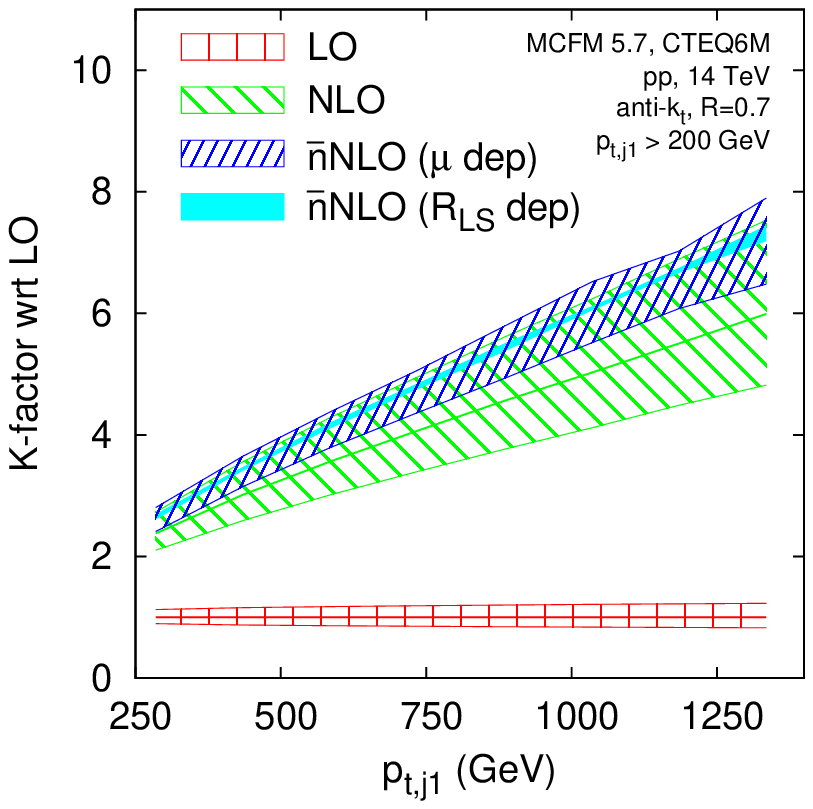}}\\
  \vspace{0.5cm}
  \scalebox{0.8}{\includegraphics{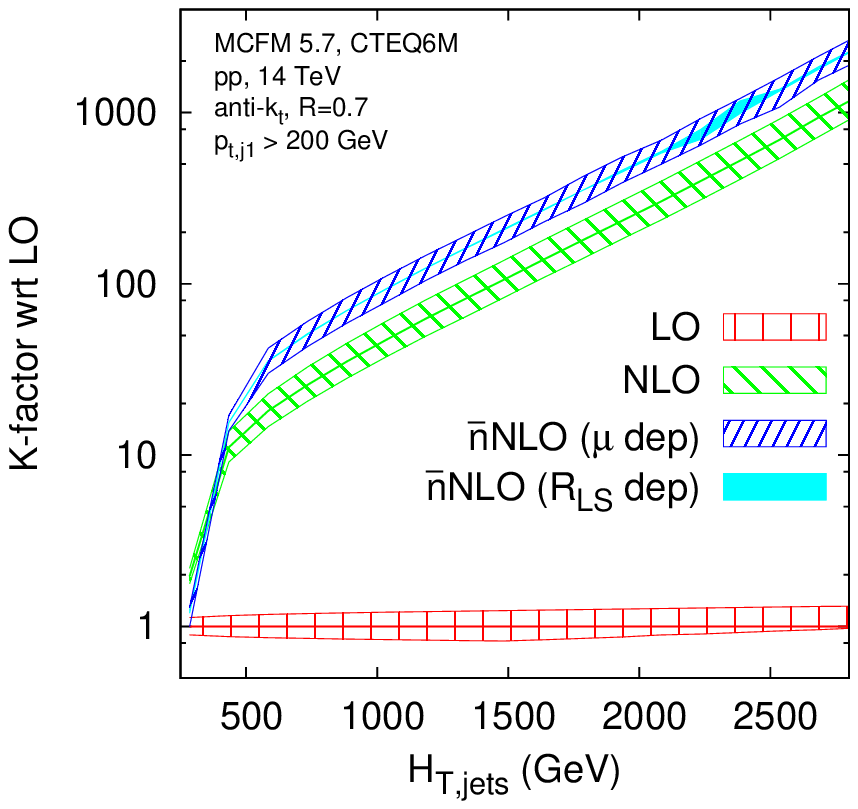}}\;\;
  \scalebox{0.8}{\includegraphics{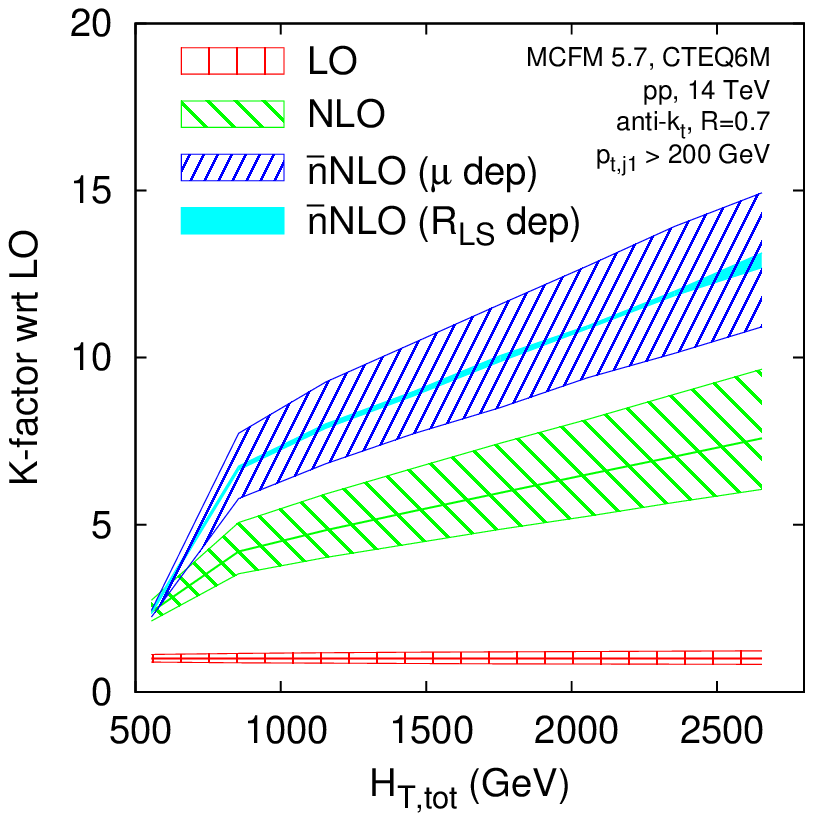}}
  \caption{Comparison of the \nNLO/LO $K$-factor with the NLO/LO
    $K$-factor, together with their scale and $R_{\LS}$ uncertainties 
    for four observables in the Z+jet process}
  \label{fig:nNLO_NLO}
\end{figure}

Results at \nNLO are given in fig.~\ref{fig:nNLO_NLO}.
In the case of $p_{t,\Z}$ the result is similar to the NLO result,
and the scale uncertainties remain largely unchanged.
In other words, since Z+2j topologies do not dominate the high-$p_{t,\Z}$
distribution, adding NLO corrections to them (i.e. \nNLO Z+j) makes
no difference either to the result or to the uncertainties. 
We have also shown the dependence on the choice of $R$ in the LoopSim
procedure.  It is smaller than the scale dependence.

The $p_{t,j1}$ distribution gets a correction that is just within
the NLO uncertainty band, with
\nNLO uncertainties that are about half the
size of the NLO band. Adding in the \nNLO term has made a real
difference. 
This is precisely what we expect: the observable is dominated by
Z+2-parton configurations, and these were only present at tree-level
in the NLO Z+j calculation.  Our use of \nNLO provides the additional
1-loop Z+2-parton and tree-level Z+3-parton configurations that come
with NLO Z+2j accuracy.

Given the improvement in scale uncertainty, we need to ask whether
the uncertainty due to $R_{\LS}$ variation might somehow eliminate
part of this benefit.
It is, however, small. The reasons are similar to those given around
eq.~(\ref{eq:RLS-uncert}).

The $H_{T,\jets}$ and $H_{T,\tot}$ distributions get significant
\nNLO corrections, with \nNLO/NLO $K$-factors of about $1.7-2$.
Absolute scale uncertainties increase slightly compared to NLO, but
because of the large $K$-factor, relative scale uncertainties
diminish.
At first sight, it is somewhat disturbing that the \nNLO and NLO uncertainty bands
don't overlap. 
Given the novelty of the LoopSim method, one should therefore ask
whether this is reasonable and whether there is any way of
cross-checking the result.

A first observation is that since \nNLO Z+j is really NLO of the
dominant Z+2j component, the 
large \nNLO corrections that we see are comparable to an $\order{2}$
$K$-factor for going from LO to NLO in the Z+2j prediction. There are
many contexts where NLO and LO results are not compatible within
scale uncertainties, and so it is not unreasonable that the same should be
seen here.

\begin{figure}
  \centering
  \scalebox{0.8}{\includegraphics{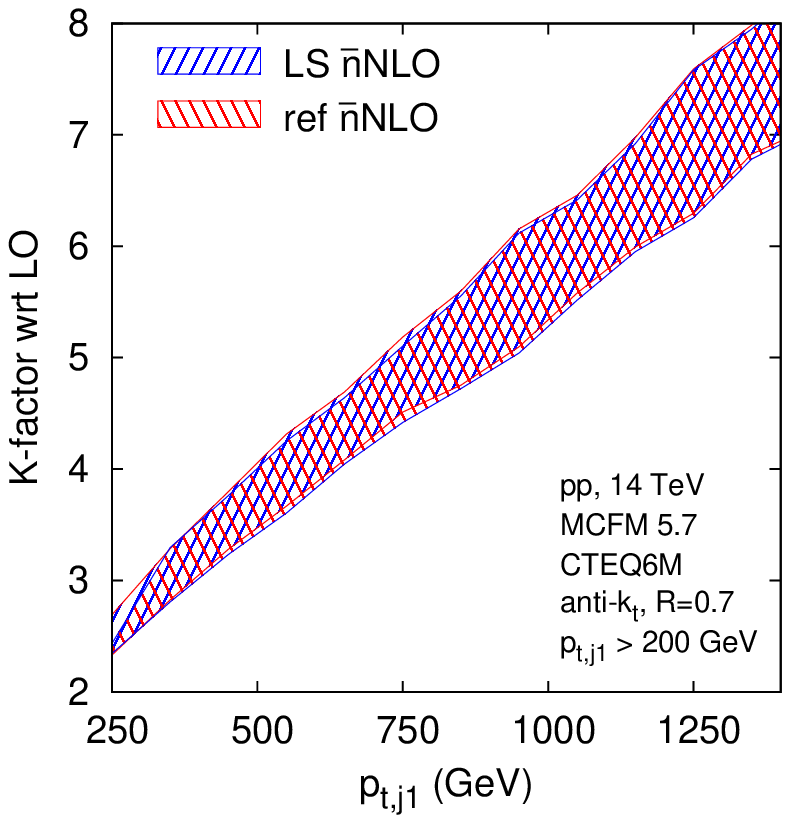}}\;\;
  \hspace{20pt}
  \scalebox{0.8}{\includegraphics{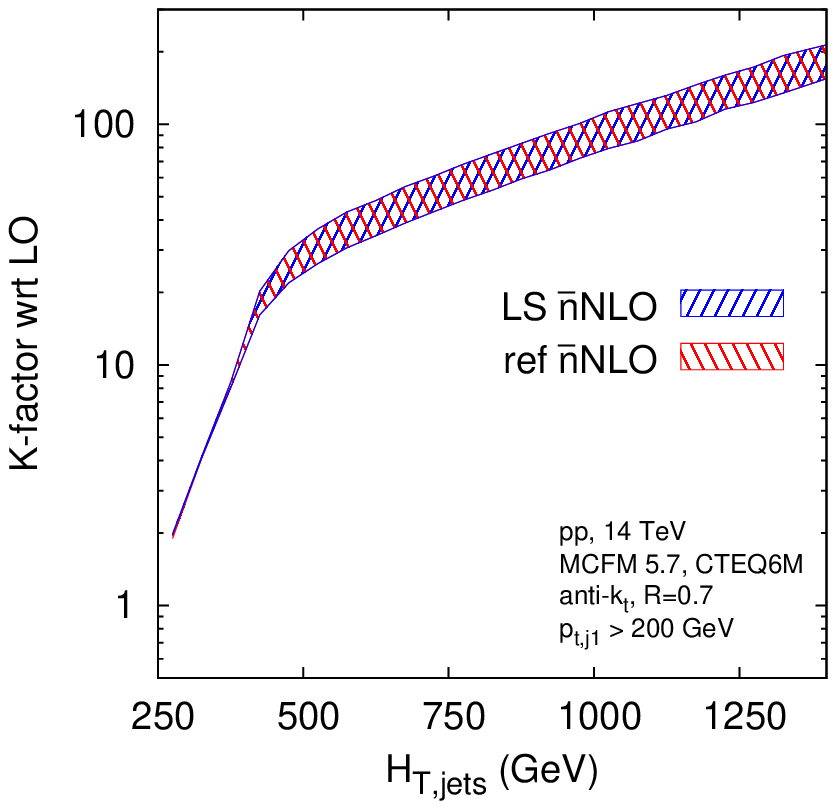}}
  \caption{Comparison between the approximate NNLO/LO $K$-factor
    calculated using respectively the LoopSim and the
    ``reference-observable'' method for $p_{t,j1}$ and
    $H_{T,\jets}$. As a reference observable we have used the differential
    cross section for $p_{t,\Z}$. }
  \label{fig:Z+j_nNLO_alt}
\end{figure}

Still, we would like to have some more quantitative cross checks that
our results are sensible. One
option is to consider the alternative ``reference-observable'' method
presented in section~\ref{sec:alternative_method}, which only makes
use of standard NLO calculations to compute the approximate NNLO
corrections.
The comparison between the two methods is shown in
fig.~\ref{fig:Z+j_nNLO_alt} for $H_{T,\jets}$ and $p_{t,j1}$, where we
have taken $p_{t,\Z}$ as the reference observable.
One notices near perfect agreement for $H_{T,\jets}$ and very good
agreement for $p_{t,j1}$.
This gives us some degree of confidence that the \nNLO LoopSim results
provide an accurate description of the NNLO behaviour for these
observables.

A second option for cross-checking the large \nNLO effects for
$H_{T,\jets}$ and $H_{T,\tot}$, is to examine whether
$H_T$ type observables might generally be ``difficult''. 
To do so we look at them in the case of QCD jet events.

\section{QCD jet events as a testing ground}
\label{sec:QCD}

We have seen that the \nNLO $K$-factors for the two effective-mass
variables, $H_{T,\tot}$ and $H_{T,\jets}$, in $\Z+$jet(s) events are
about a factor of two above the NLO $K$-factor.

Since NLO is the first order at which we see the dominant ``dijet''
topology for the $H_{T}$ variables in Z+jet(s),
fig.~\ref{fig:z+jet-diags}B,C, it might be 
instructive to establish a correspondence with a simpler process, QCD
dijet production.
Having a NLO Z+j prediction is analogous to a LO dijet prediction; and
the \nNLO Z+j predictions should be analogous to NLO dijet
predictions.
NLO cross sections for dijet observables can be calculated exactly and
therefore we can check whether NLO $K$-factors of order $2$ appear for
effective-mass observables in pure QCD events.

We will consider several effective-mass observables: an $H_{T,n}$
variable, which sums over the $n$ hardest jets above some threshold
($p_{t,\min} = 40\GeV$; such a cut is often imposed
experimentally\footnote{In section~\ref{sec:results} we did not apply
  this kind of cut on the $H_T$ variables; one purpose in applying it
  here is to ascertain whether the large higher-order effects persist
  even with it.})
\begin{equation}
  \label{eq:HTn}
  H_{T,n} = 
  \hspace{-1.5em}
  \sum_{i \in \text{jets with $p_{t,ji}>p_{t,\min}$}}^n 
  \hspace{-2.5em}
  p_{t,ji}\,,
\end{equation}
where $p_{t,i}$ is the transverse momentum of the $i^\text{th}$
hardest jet.
Upper limits on the number of jets included in the effective mass are
common in SUSY searches \cite{Aad:2009wy,Ball:2007zza}.
We also define an effective mass for all jets above the $p_{t,\min}$
threshold,
\begin{equation}
  \label{eq:HT-all}
  H_{T} \equiv H_{T,\infty}\,,
\end{equation}
which is similar to the $H_{T,\jets}$ and $H_{T,\tot}$
observables of section~\ref{sec:results}.
Finally, for completeness we will consider the distributions of $p_{t,j1}$,
$p_{t,j2}$ and the inclusive jet spectrum.
All our results in this section will be for a centre-of-mass energy of
$7\TeV$, to allow comparison to results in the current run of the LHC.

At LO, the distributions of $\frac12 H_{T,n}$ ($n\ge 2$), $\frac12
H_{T}$, $p_{t,j1}$, and $p_{t,j2}$ will all be identical.
The inclusive jet spectrum will have a distribution that is twice as
large (because each of the two jets contributes).
Note that we do not impose any rapidity acceptance limits on the jets:
though such a cut would have been trivial to include in the LoopSim
procedure, it would have complicated somewhat the reference-observable
approach that we will consider at the end of the section.
LoopSim results with a rapidity cuts of $|y|<2$ on the jets are available from
the authors on request.

\begin{figure}[t]
  \centering
  \includegraphics[width=0.47\textwidth]{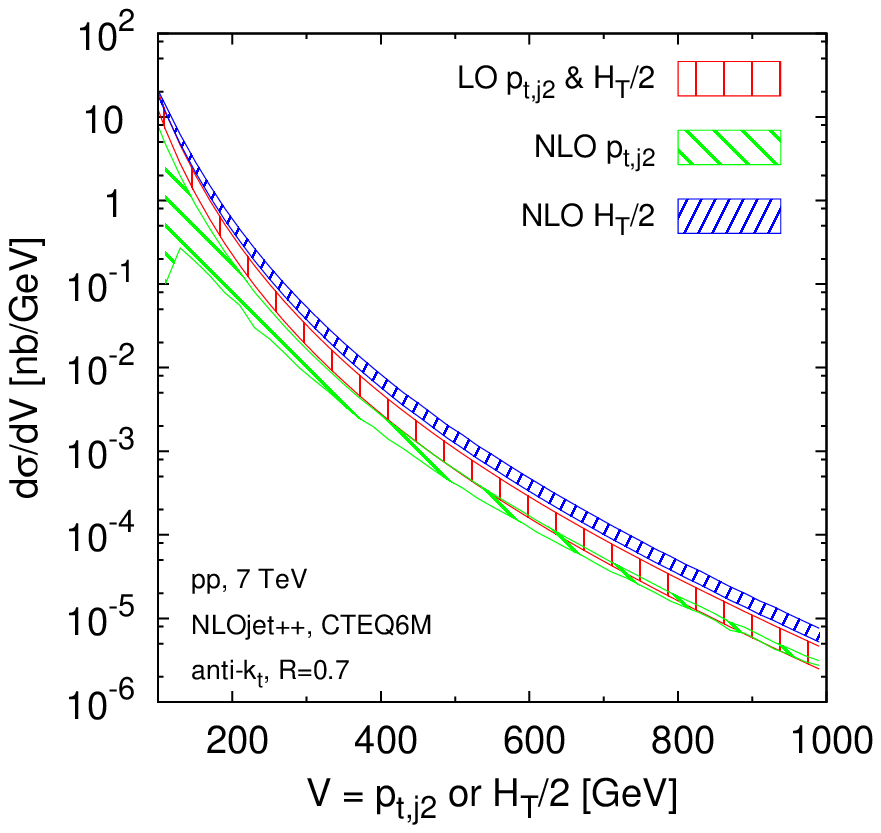}\quad
  \includegraphics[width=0.38\textwidth]{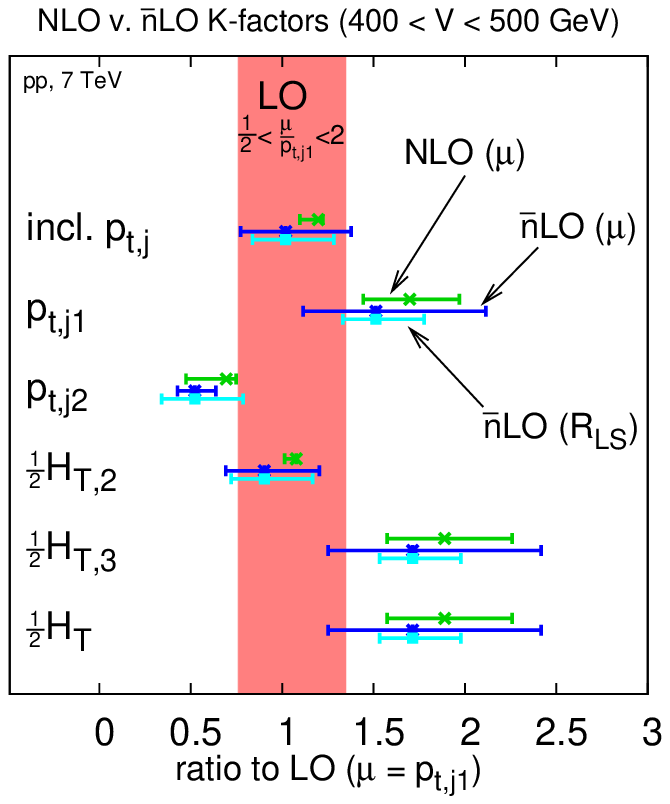}
  \caption{Left: differential cross sections for the $p_{t,j2}$ 
     and
    $\frac12 H_T$ observables, at LO, where they are identical, and at
    NLO where they have substantially different $K$-factors.
    Right: the NLO $K$-factors for the $400 < V/\GeV < 500$ bin for each
    choice of variable $V$ among the following: the inclusive jet
    spectrum, the $p_t$ distribution of the hardest ($p_{t,j1}$) and
    second  hardest ($p_{t,j2}$) jets, (half) the effective mass of the two
    hardest jets ($H_{T,2}$), three hardest jets ($H_{T,3}$) and of
    all jets above $40\GeV$ ($H_{T}$).
    Also shown on the right are the \nLO results for the $K$-factors.
    The NLO and \nLO($\mu$) widths correspond to the
    uncertainty due to
    simultaneous renormalisation and factorisation scale variation
    by a factor of two around a central value $\mu=p_{t,j1}$. 
    The \nLO\!\!($R_\LS$) width shows the uncertainty from a variation of
    $R_{\LS}$ in the range $0.5 < R_{\LS} < 1.5$.
  }
  \label{fig:jet-dist+nlo-kfacts}
\end{figure}

Figure~\ref{fig:jet-dist+nlo-kfacts}(left) shows the distributions for
two observables, $\frac12 H_T$ and $p_{t,2}$ at LO (where they are
identical) and at NLO, as determined using
NLOJet\texttt{++}~\cite{Nagy:2001fj,Nagy:2003tz} with 
CTEQ6M PDFs.
A first comment is that $H_T$ receives a NLO $K$-factor of order $2$,
just like the \nNLO enhancements in the Z+j case.
This provides supporting evidence as to their legitimacy.
A second comment is that the cross sections are large: these
observables will be easily accessible with a few pb$^{-1}$ of
integrated luminosity at a $7\TeV$ LHC, allowing for an early
experimental verification of the large $K$-factor for $H_T$.
The other observable in the left-hand plot of
fig.~\ref{fig:jet-dist+nlo-kfacts}, $p_{t,j2}$, has a very different
$K$-factor, somewhat below $1$.
The right-hand plot shows the NLO $K$-factors for our full range of
observables, focusing on a single bin of the left-hand one, from
$400-500\GeV$.
The pattern that we see here allows us to make some deductions.
Firstly, the $H_{T,2}$ variable, which sums the $p_t$'s of the two
leading jets, is free of large NLO enhancements. 
It is the addition of the third jet in  $H_{T,3}$ and  $H_{T}$ that
brings about the enhancement. 
A natural interpretation is the following: it is common for a third,
soft jet to be present due to initial state radiation.
This third jet shifts the $H_{T}$ distribution to slightly larger
values, and because the distribution falls very steeply, that leads to
a non-negligible enhancement. This suggests that if, in
section~\ref{sec:results}, we had used effective mass observables with
at most two objects in the sum, then the \nNLO/NLO ratios would have
been close to $1$. We have verified that this is indeed the case.

The pattern for $p_{t,1}$ and $p_{t,2}$ in
fig.~\ref{fig:jet-dist+nlo-kfacts} can also be explained in
similar terms: a soft ISR emission boosts the hard dijet system,
breaking the degeneracy between the $p_t$'s of the two hardest
jets. It is jet $1$ that shifts to larger $p_t$ (giving a $K$-factor
$> 1$), while jet $2$ shifts to lower $p_t$ and so it gets a $K$
factor below $1$.
For the inclusive jet spectrum, and for $H_{T,2}$, this effect
balances out.
In addition, final-state radiation from one of the jets can cause it
to shift to lower $p_t$ (becoming the 2nd jet), further reducing the
$K$-factor for the distribution of $p_{t,j2}$.

Of the different variables, it is only the inclusive jet $p_t$ and
$H_{T,2}$ for which there is a clear reduction in scale uncertainty in
going from LO to NLO.

Figure~\ref{fig:jet-dist+nlo-kfacts}(right) also shows the \nLO
results (including uncertainties both from scale variation and from
the LoopSim parameter $R_{\LS}$).
Despite the fact that none of the $K$-factors is parametrically large
(except arguably for $H_{T,3}$ and $H_{T}$), the \nLO results are
remarkably effective at reproducing the pattern of NLO $K$-factors,
albeit with a small systematic shift and generally larger scale
uncertainties.
One can also verify that, to within $10-20\%$, the $p_t$ dependence of
the NLO $K$-factors is reproduced at \nLO.

\begin{figure}
  \centering
  \includegraphics[width=\textwidth]{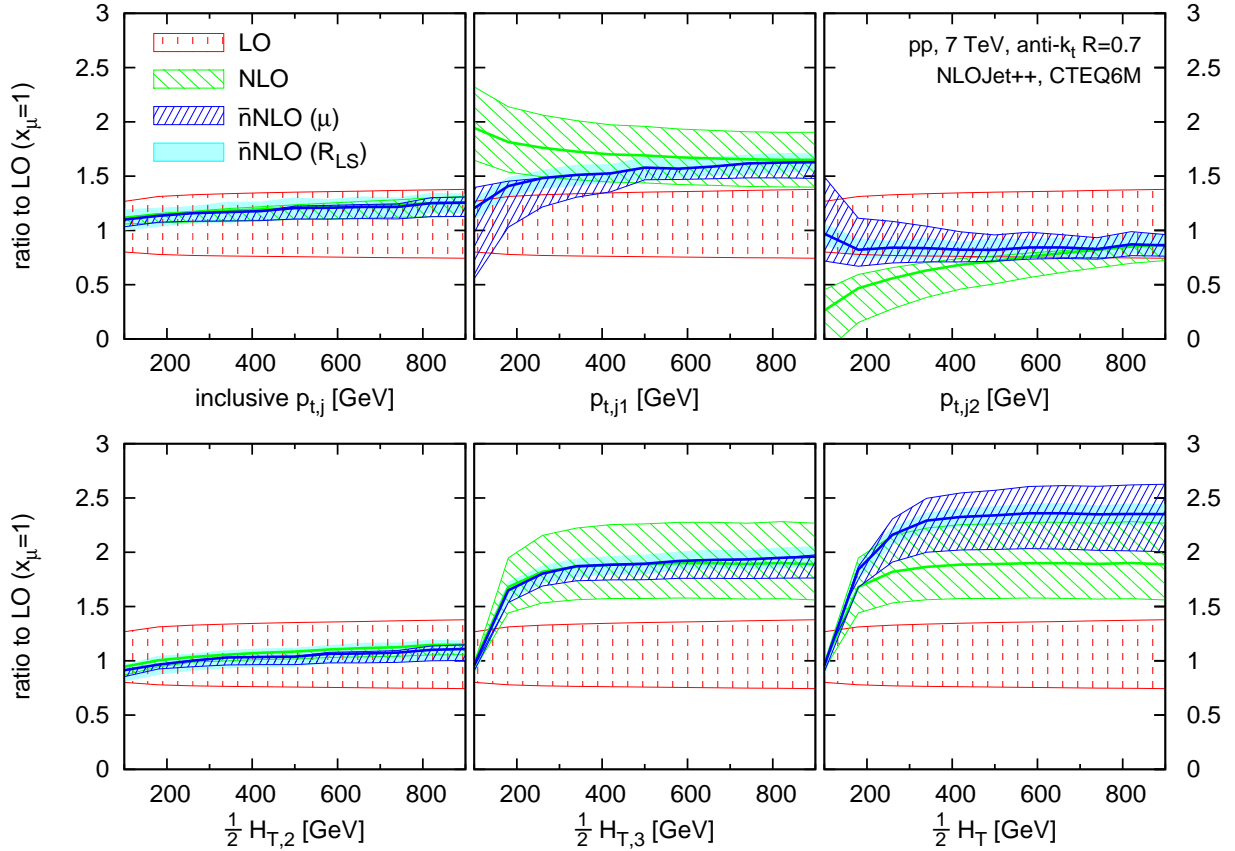}
  \caption{The \nNLO and NLO $K$-factors relative to the LO
    predictions, as a function of $p_t$ (or $\frac12 H_T$, etc.), for
    the collection of jet observables described in the text. }
  \label{fig:jets-K-nNLO}
\end{figure}

Given this success of \nLO, and the observed limited convergence of
some of the observables at NLO, it is interesting to examine what
happens at \nNLO, where the additional 3j@NLO contribution that we
require is again obtained using NLOJet\texttt{++}.
Results are shown in fig.~\ref{fig:jets-K-nNLO}.

For the inclusive jet spectrum and $H_{T,2}$, which already saw large
reductions in scale-dependence at NLO, the \nNLO corrections have
essentially no meaningful effect: they neither significantly affect
the central values, nor reduce the scale uncertainties.
For these observables, NLO already converged well, and adding a subset
of the NNLO corrections without the 2-loop part cannot improve the
result. 

For the other effective mass observables, the situation is quite
different.
With $H_{T,3}$, the \nNLO result is close to the NLO result and the
scale uncertainty is much reduced, i.e.\ this observable seems to come
under control at \nNLO. 
In contrast, $H_{T}$ is subject to quite a large further correction, with the
central value at \nNLO lying outside the NLO uncertainty band, and the
\nNLO uncertainty band (dominated by scale variation) only marginally
smaller than at NLO.
Why is this? Perhaps we are seeing the effect of a second ISR
emission, which shifts the $H_T$ distribution to even higher
values? 
Given that $H_{T,3}$ converges and $H_T$ does not, such an explanation
is not unattractive. It is also consistent with the decrease in
$K$-factor at low $H_{T}$, where the $40\GeV$ $p_t$ cutoff on the jets
contributing to the $H_{T}$ sum will eliminate the ISR enhancement.
A definitive conclusion would however probably require further study.

For the remaining two observables, $p_{t,1}$ and $p_{t,2}$, the \nNLO
contribution goes in the opposite direction from the NLO correction
and at low $p_t$ it seems that the series fails to converge.
This is, we believe, closely related to observations of
insufficiencies of NLO predictions for dijet cross sections in DIS and
photoproduction when identical $p_t$ cuts are imposed on both jets
\cite{KK,symm-cuts,Chekanov:2001fw,Aktas:2003ja,Banfi:2003jj}
(equivalent to integrating the $p_{t2}$ distribution above that cut).
The worse convergence at low $p_t$ is probably due to the larger fraction
of subprocesses that involve gluons in the underlying $2\to2$ scattering,
so that perturbative corrections tend to go as $(C_A \as/\pi)^n$ rather
than as $(C_F \as/\pi)^n$ at higher $p_t$.

Considering that we do not have giant NLO $K$-factors for the jet processes
shown here, one may question the validity of the information obtained
from the LoopSim procedure.
An important cross check comes from a comparison with the
reference-observable technique. 
Examining fig.~\ref{fig:jet-dist+nlo-kfacts} (right), one sees two
natural reference observables: the inclusive jet spectrum and
$H_{T,2}$, both of which show ``perturbative'' $K$-factors and small
scale dependence at NLO.
Here we will use (half) the inclusive jet spectrum as the reference
observable (results with $H_{T,2}$ would be almost identical).

\begin{figure}
  \centering
  \includegraphics[width=0.38\textwidth]{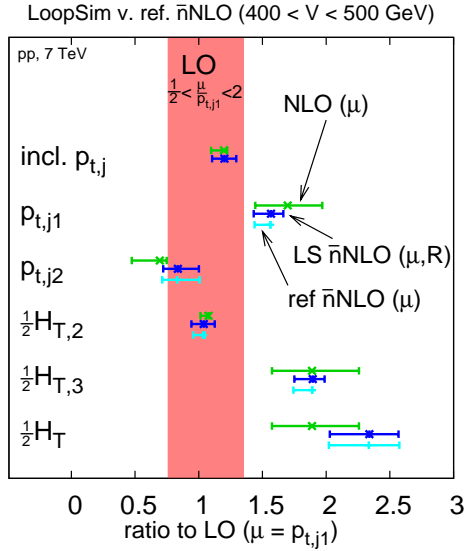}
  \caption{Comparison of LoopSim based \nNLO results with those from
    the reference-observable method, here using the inclusive jet
    $p_t$ spectrum as the reference observable. In the LoopSim results
    (labelled \nNLO), the uncertainty bar spans the envelope of the
    scale and $R_{\LS}$ uncertainties. The results are for the $400 <
    V/\GeV < 500$ bin for each observable $V$, as in
    fig.~\ref{fig:jet-dist+nlo-kfacts}. }
  \label{fig:jets-nnlo-ref}
\end{figure}

Figure~\ref{fig:jets-nnlo-ref} provides a comparison of the LoopSim
\nNLO results (showing the envelope of the scale and $R_{\LS}$
uncertainties) with the reference-observable \nNLO results. 
The comparison is given for all observables except the reference
observable itself. 
The agreement between the two methods is striking, with the
reference-observable method giving just a small shift of the
$K$-factors relative to the LoopSim results. The shift is identical
for all the observables, as it has to be: it is simply equal to the
difference between the NLO and \nNLO results for the reference
observable.
Insofar as we believe the scale dependence to be representative of
the true NLO uncertainty on the inclusive jet spectrum,\footnote{In
  light of the
  fact that the \nNLO uncertainty for the inclusive jet spectrum is
  larger than the NLO uncertainty, it may be that our symmetric scale
  variation is underestimating somewhat the uncertainties
  present at NLO.
  To be conservative, it might have been safer to vary the
  renormalisation and factorisation scales independently.}
the results for the other observables should therefore be good
approximations to the full NNLO results.

\section{Conclusions}
\label{sec:concl}

Several cases of LHC observables with giant NLO $K$-factors have come
to light in recent years.
They are characterised by the presence at NLO of new partonic
scattering topologies that have large enhancements over the LO
topologies.
In these cases, NLO calculations, while important in highlighting the
presence of the large $K$-factors, cannot on their own provide
accurate predictions.

In this article we have examined how to address this problem by combining
NLO results for different multiplicities, for example Z+j@NLO with
Z+2j@NLO.
Our main, most flexible method, LoopSim, makes use of unitarity to
cancel the infrared and collinear divergences that appear when one
tries, say, to apply Z+2j@NLO calculations to observables that are
non-zero starting from Z+1-parton.
We referred to the result as Z+j@\nNLO, where the ``\nbar'' indicates
that the highest loop contribution to the NNLO result (the two-loop
part) has been estimated with LoopSim.

In introducing a new approximate method for estimating NNLO
corrections, significant evidence needs to be provided that the method
is meaningful.
Firstly, we gave reasons why, in cases with giant $K$-factors associated
with new NLO topologies, we expect \nNLO results to be a good
approximation to NNLO results.
As a next step, we carried out studies comparing
Z/$\gamma^*$@\nNLO (DY) to NNLO predictions for the $pp\to
Z/\gamma^*+X\to e^+e^-+X$ process.
In comparing the DY lepton $p_t$ \nNLO distributions to NNLO we found
near-perfect agreement in a region of giant $K$-factors, $p_t -
\frac12 m_\Z \gtrsim \Gamma_Z$.
Interestingly, even in the region where the NLO $K$-factor was not
large, $p_t\lesssim \frac12m_Z$, the \nNLO results provided a
significantly better approximation 
to NNLO than did the plain NLO result.
This need not always be the case, but is, we believe, connected to the
observation that our \nLO results reproduced much of the structure
seen at NLO (recall, Z@\nLO means combining Z@LO with Z+j@LO).

For Z+j production, the first step of our validation procedure was to
compare \nLO and NLO results.  All observables with giant $K$-factors
showed good agreement between the two (one with a moderately large
$K$-factor did not).
For those observables, \nNLO always appeared to provide extra
information: either suggesting a convergence of the perturbative
series, with reduced scale uncertainties (for $p_{t,j1}$), or an
indication of substantial further higher order corrections (for the
effective-mass type observables $H_{T,\jets}$ and $H_{T,\tot}$).
Almost identical results were seen with our alternative
``reference-observable'' estimate of the NNLO contribution.

The large \nNLO corrections that we saw for effective mass observables
led us to examine a range of effective-mass and jet observables in the
simpler context of pure jet events (with the expectation that Z+j@NNLO
might be similar to 2j@NLO). There we saw a significant NLO $K$-factor
for all effective mass variables except one, $H_{T,2}$, which summed
over just the two leading jets.
In the Z+j case we had summed over all jets and hence it is not surprising
that we should have observed substantial \nNLO/NLO ratios.

Even though the observables in the pure jets case did not display
giant $K$-factors, the pattern of NLO results was remarkably well
reproduced at \nLO.
This encouraged us to study \nNLO predictions, which
provided substantial extra information for several of the
observables, with the reference-observable method again giving
important cross checks.
Since the cross sections for the jet observables are large, these
results could easily be tested with early LHC data.

We close this article with a few lines on the relation between LoopSim
and other predictive methods.
There is a close connection between \nLO (or \nnLO) and CKKW and MLM
\cite{Catani:2001cc,Alwall:2007fs} matching, since they also both
provide ways of combining tree-level results with different
multiplicities.
Of course CKKW and MLM matching provide an interface with parton
showers too, which the LoopSim method does not.
On the other hand it is significantly easier to include multiple loop
orders into the LoopSim method than it is within
matrix-element/parton-showering matching procedures (though work is
ongoing in this direction see e.g.\ \cite{Lavesson:2008ah}).

An interesting cross-check of the LoopSim method will come with the
completion of the NNLO calculations for the Z+j and dijet processes.
At that point the method could also, for example, be used to merge
Z@NNLO with Z+j@NNLO, so as to provide an \nNNLO prediction for
quantities like the Drell-Yan lepton $p_t$ spectrum.
The value of the LoopSim method also goes hand-in-hand with progress
on 1-loop calculations, especially with the prospect of automated
NLO calculations now on the horizon (for example
\cite{KeithEllis:2009bu,Berger:2009ep,Bevilacqua:2010ve}).

Note that the LoopSim code, which will be made public in due course,
can currently only deal with hadron-collider processes involving any
number of light partons and up to one vector boson. It would benefit
from further work to appropriately include heavy quarks and additional
bosons.

\section*{Acknowledgements}

We are grateful to Fabio Maltoni and Simon de Visscher for stimulating
discussions and for sharing results with us on the behaviour of MLM
type matching for the Z+jet processes, to Giulia Zanderighi for
insightful comments and a careful reading of the manuscript, to
Massimiliano Grazzini and Giancarlo Ferrera for assistance with
DYNNLO, to Matteo Cacciari and Gregory Soyez for FastJet development
work that has been useful in implenting the LoopSim procedure and to
Philippe Schwemling and Dirk Zerwas for helpful exchanges about $H_T$
definitions.
GPS wishes to thank the High Energy Theory group at Rutgers
University, the Princeton University Physics Department and the Aspen
Center for Physics for hospitality during various stages of this work.
This work was supported by the French Agence Nationale de la
Recherche, under grant ANR-09-BLAN-0060.


\appendix

\section{Recoil procedure \label{app:recoil_procedure}}

In this appendix, we provide further details on how we perform the
recoil of an event when a particle becomes virtual, including the
treatment of the decay products of the Z boson. We first examine the
simpler case of a particle that makes a loop with the beam, then we
show how to deal with a particle that makes a loop with another
particle.

\subsection{A particle recombines with the beam}

Let us assume that particle $i_0$ makes a loop with the beam. To
balance the event moment, we follow the following procedure:
\begin{enumerate}
  \item For each particle $i\neq i_0$, store its rapidity $y_i$.
  \item Perform a separate longitudinal boost on each particle so as
    to bring its rapidity to $0$ (\ie get a purely transverse event).
  \item Compute
    \begin{equation}
      E_{tot} = \sum_{i\neq i_0}E_i\,,
    \end{equation}
    where $E_i$ is the energy of particle $i$ in the purely transverse
    event.
  \item Define
    \begin{equation}
      \label{eq:pt-balance-boost}
      k = \left(E=E_{tot},\vec{p}_t = \vec{p}_{t,i_0},p_z = 0\right)\,,
    \end{equation}
    and boost all particles into the rest frame of $k$ (so that the
    total transverse momentum balances).
  \item Perform a longitudinal boost on each particle so that it
    recovers its original rapidity $y_i$.
\end{enumerate}
For the case where two particles, $i_0$ and $i_1$, are looped with the
beam, replace $i\neq i_0$ with $i\neq i_0,i_1$ and in
eq.~(\ref{eq:pt-balance-boost}) replace $\vec{p}_{t,i_0}$ with
$\vec{p}_{t,i_0} + \vec{p}_{t,i_1}$, etc.
In the case where the Z decays, for instance into $2$ leptons, the
procedure is identical except that we apply to the leptons the same
longitudinal boosts as for the Z (the rapidity of the leptons is
thus not necessarily $0$ when we apply the transverse boost). This 
conserves the property that the sum of the leptons' momenta is
still the Z momentum in the ``looped'' event.

The logic of the above procedure is that if we had attempted to apply
a transverse boost without stages 2 and 5, we would have found that
our choice of transverse boost, and the corresponding mapping of
high-$p_t$ particles' momenta, would be affected by the presence of
energetic particles collinear to the beam. This would have made the
procedure collinear unsafe.

\subsection{A particle recombines with another particle}

\begin{figure}[t]
  \centering
  \includegraphics[scale=0.3]{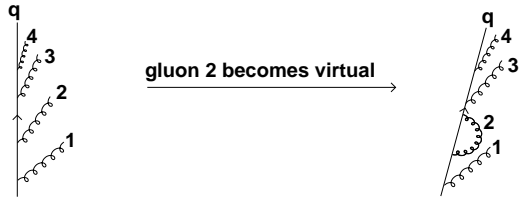}
  \caption{Case where four gluons are emitted from the same
    quark. Gluon $1$ is the last to be clustered with the quark (which
    roughly corresponds to an early time emission) and gluon $4$ is
    the first to be clustered. In the event where gluon $2$ makes a
    loop over the quark, we spread the gluon $2$'s momentum over the
    quark's momentum and the momenta of gluons that were emitted {\it
      after} it, \ie gluons $3$ and $4$ (an earlier time emission like
    gluon $1$ cannot be affected).}
  \label{fig:pp-recombination}
\end{figure}

Let us consider the situation depicted in
fig.~\ref{fig:pp-recombination}: four gluons are emitted from the same
quark, but at different angles:
\begin{equation}
  \theta_{1q}\gg\theta_{2q}\gg\theta_{3q}\gg\theta_{4q}\,,
\end{equation}
and gluon $2$ becomes virtual. The virtualisation of gluon $2$ over
the quark cannot have an impact on gluon $1$, which was emitted
earlier in an angular-ordered picture. But it has an impact on gluons
$3$ and $4$. More precisely, let the $p_i$ be the momenta in the
original event and $p'_i$ the momenta in the event where gluon $2$ is
virtual.
We define
\begin{equation}
  \label{eq:reshuffle-looped-momenta-tot}
  p_{t,\tot} = p_{t,q}+p_{t,3}+p_{t,4}\,.
\end{equation}
and then set the $p_i'$ as follows:
\begin{subequations}
  \begin{align}
    \label{eq:reshuffle-looped-momenta}
    p'_i & = p_i + \frac{p_{t,i}}{p_{t,\tot}}p_2\quad\mbox{ for $i=$ $q$, $3$, $4$}\,,\\
    p'_1 & = p_1\,.
  \end{align}
\end{subequations}
Subsequently each particle's $p_i'$ momentum is adjusted such that its mass
is $0$ (or $m_\Z$ if gluon $3$ is a Z boson rather than a gluon),
keeping its transverse components $p_x$, $p_y$ and its rapidity
unchanged.
This can be easily generalised to any number of particles recombining
with the same hard one: for each recombined particle $i$, we spread
the looped particle over the hard particle $h$ and over any non-looped
emissions from $h$ that are at smaller angle (i.e.\ earlier in the
C/A clustering sequence) than $i$. In
eqs.~(\ref{eq:reshuffle-looped-momenta-tot},\ref{eq:reshuffle-looped-momenta})
it is always the original particle momenta that are used to determine
the $p_{t,i}/p_{t,\tot}$ ratio, so that the result is independent of
the order in which we perform the recombinations.

This procedure is designed to ensure collinear safety: if, for instance, 
gluon $4$ is collinear to the quark in the original event, then it remains
collinear in the looped event. 
And if it is gluon $4$ (emitted after all the others) that is looped,
only the quark momentum is rescaled and its direction barely changes,
so that angles between the quark and the other gluons stays the same.

In the case where the Z decays into $2$ leptons, one applies the
following procedure to each of the leptons:
\begin{enumerate}
\item Perform a longitudinal boost of the Z boson respectively in
  the original event and the looped event such that it has $0$
  rapidity in each case. Call the momenta obtained
  $p_{Z,0}=(E_0,\vec{p}_{t,0},0)$ and $p_{Z,1}=(E_1,\vec{p}_{t,1},0)$
  respectively.
\item Perform a longitudinal boost of the lepton from the original
  event into the frame where the initial Z has $0$ rapidity.
\item Define a purely transverse vector $k$ such that $p_{Z,0}$ is
  transformed to $p_{Z,1}$ if it is boosted into $k$'s rest frame:
    \begin{equation}
      k = \left(E_1+E_0,\frac{2}{1+C}(\vec{p}_{t,1}-\vec{p}_{t,0}),0\right)\,,
    \end{equation}
    with
    \begin{equation}
      C = \frac{(\vec{p}_{t,1}-\vec{p}_{t,0})^2}{(E_1+E_0)^2}\,.
    \end{equation}
  \item Boost the lepton's momentum into $k$'s rest frame.
  \item Apply to the lepton the longitudinal boost that brings
    $p_{Z,1}$ to its true rapidity in the looped event.
\end{enumerate}

We are aware of the cumbersome nature of these procedures. A
simplification of them that retained the relevant collinear-safety
properties would certainly be of interest.

\section{The LoopSim method and incoming partons}
\label{sec:incoming-partons}

Without going into a full proof, we shall here illustrate why the
LoopSim method is sensible even in the presence of incoming hadrons,
by considering what happens at \nLO.
We start with a LO cross section for a process producing $n$ hard
objects 
\begin{equation}
  \label{eq:pdf-LO-start}
  \sigma_n^{\LO} = \int dx_a dx_b \, d\Phi_n \, \frac{d \hat \sigma_{ij \to
      n}(x_a
  p_a, x_b p_b)}{d\Phi_n} \, f_{i/a}(x_a, \muF^2) f_{j/b}(x_b, \muF^2)\,
  C (p_1,\ldots, p_n)\,.
\end{equation}
For compactness of notation, we have dropped the $\muR$ dependence in
the differential tree-level partonic cross section $d\hat \sigma_{ij
  \to n}/d\Phi_n$. We have also not yet specified our choice for the
factorisation scale $\muF$.
We assume that $d\hat \sigma_{ij \to n}/d\Phi_n$ contains the
necessary constraints to relate the incoming partonic momenta to the
outgoing momenta.
We further integrate over the phase-space $dx_a dx_b d\Phi_n$, and
include a function $C (p_1,\ldots, p_n)$, which is $1$ if the momenta
pass our cuts and $0$ otherwise.

We now imagine that there is some transverse-momentum scale $Q_0$
below which no radiation is emitted. To $\order{\as}$, the PDFs
$f_{i/a}(x_a, \muF^2)$ can be written in terms of PDFs at scale
$Q_0$:
\begin{equation}
  \label{eq:pdf-from-Q0}
  f_{i/a}(x_a, \muF^2) = f_{i/a}(x_a, Q_0^2) + 
  \frac{\as}{2\pi} \int_{Q_0^2}^{\muF^2}
  \frac{dk_t^2}{k_t^2} \int \frac{dz}{z}
  P_{ik}(z)
  f_{k/a}(x/z, Q_0^2)\,,
\end{equation}
where we sum implicitly over repeated indices.
We also define an unregularised splitting function $p_{ik}(z)$
such that $P_{ik}(z) = p_{ik}(z) - \delta(1-z) \int dz' \bar
p_{ik}(z')$, with $\bar p_{ik}(z')$ embodying the virtual parts of the
splitting function (it is zero for $i \neq k$).

Next, we write the LO cross section in terms of a PDF for proton $a$
that has been evaluated at scale $Q_0$:
\begin{multline}
  \label{eq:pdf-LO-Q0}
  \sigma_n^{\LO} = \int dx_a dx_b \, d\Phi_n \, \frac{d \hat \sigma_{ij \to
      n}(x_a
  p_a, x_b p_b)}{d\Phi_n} \,  f_{j/b}(x_b, \muF^2)\,
  C (p_1,\ldots, p_n)
  \\
  \times \left[
    f_{i/a}(x_a, Q_0^2)
    + 
    \frac{\as}{2\pi} \int_{Q_0^2}^{\muF^2} \frac{dk_t^2}{k_t^2} dz\, 
    \left(
      \frac{p_{ik}(z)}{z} f_{k/a}(x_a/z, Q_0^2)
      -
       \bar p_{ik}(z) f_{i/a}(x_a, Q_0^2)
    \right)
  \right]\,.
\end{multline}
Note that the first term in round brackets on the second line
corresponds to real emission of a parton. However that parton is not
taken into account in the $C(p_1,\ldots, p_n)$ factor.

Next we examine the structure of the \nLO contribution,
\begin{multline}
  \label{eq:pdf-nLO-start}
  \sigma_n^{\nLO} = \sigma_n^{\LO} + \int dx_a dx_b \, d\Phi_{n+1} \, 
  \frac{d \hat \sigma_{ij \to {n+1}}(x_a p_a, x_b p_b)}{d\Phi_{n+1}} 
  \, f_{i/a}(x_a, \muF^2) f_{j/b}(x_b, \muF^2)\,
  \\
  \times
  \left[
    C (p_1,\ldots, p_{n+1})
    - 
    C (p_1^\LS,\ldots, p_{n}^\LS)
  \right]\,,
\end{multline}
where the $p_1^\LS\ldots p_n^\LS $ represent the momenta when the
LoopSim procedure has looped $p_{n+1}$.
In the limit in which $p_{n+1}$ is collinear to incoming parton $i$,
with momentum $p_{n+1} \simeq (1-z)x_a p_a$, the $n\!+\!1$-parton differential
cross section and phase-space simplify
\begin{equation}
  \label{eq:sigma-splitting}
  dx_a
  d\Phi_{n+1}  
  \frac{d \hat \sigma_{ij \to {n+1}}(x_a p_a, x_b p_b)}{d\Phi_{n+1}} 
  = 
  dx_a'
  d\Phi_{n}
  \frac{d \hat \sigma_{kj \to {n}}(x_a' p_a, x_b p_b)}{d\Phi_{n}}  
  \cdot
  \frac{\as}{2\pi} \frac{dz}{z} \frac{dk_{t,n+1}^2}{k_{t,n+1}^2} p_{ki}(z)\,,
\end{equation}
where $x_a' = z x_a$. By ``collinear'' we will mean $k_{t,n+1} \ll Q$
where $Q$ is the momentum transfer in the hard process.
In this limit we also have that $p_l^\LS \simeq p_{l}$ (for $l \le n$).
So, still working within the collinear limit, we can now rewrite
eq.~(\ref{eq:pdf-nLO-start}) as
\begin{multline}
  \sigma_n^{\nLO} \simeq \sigma_n^{\LO} + \int dx_a' dx_b \, d\Phi_{n} \, 
  \frac{d \hat \sigma_{kj \to {n}}(x_a' p_a, x_b p_b)}{d\Phi_{n}} 
  \,  f_{j/b}(x_b, \muF^2)\,
  \\
  \times
  \frac{\as}{2\pi} \int_{Q_0^2}^{Q^2} \frac{dk_{t,n+1}^2}{k_{t,n+1}^2}
  \frac{dz}{z}\, p_{ki}(z) 
  \left[
    C (p_1,\ldots, p_{n+1})
    - 
    C (p_1,\ldots, p_{n})
  \right]
  f_{i/a}(x_a'/z, \muF^2)\,.
\end{multline}
Next, we exchange $i \leftrightarrow k$, replace $x_a' \to x_a$ and
change the scale $\muF^2$ in $f_{i/a}(x_a'/z, \muF^2)$ to be $Q_0^2$,
which is allowed because it corresponds to an $\order{\as^2}$ change
(while here we consider only $\order{\as}$):
\begin{multline}
  \label{eq:pdf-nLO-end}
  \sigma_n^{\nLO} \simeq \sigma_n^{\LO} + \int dx_a dx_b \, d\Phi_{n} \, 
  \frac{d \hat \sigma_{ij \to {n}}(x_a p_a, x_b p_b)}{d\Phi_{n}} 
  \, f_{j/b}(x_b, \muF^2)\,
  \\
  \times
  \frac{\as}{2\pi} \int_{Q_0^2}^{Q^2} \frac{dk_{t,n+1}^2}{k_{t,n+1}^2}
  \frac{dz}{z}\, p_{ik}(z) 
  \left[
    C (p_1,\ldots, p_{n+1})
    - 
    C (p_1,\ldots, p_{n})
  \right]
  f_{i/a}(x_a/z, Q_0^2)\,.
\end{multline}
Note now that if we take $\muF^2 \sim Q^2$ in eq.~(\ref{eq:pdf-LO-Q0}),
then the second term in
square brackets in eq.~(\ref{eq:pdf-nLO-end}) cancels the first term
in round brackets in the second line of eq.~(\ref{eq:pdf-LO-Q0}).
In other words for initial-state radiation, the action of LoopSim is
not so much to provide virtual corrections as to cancel the
real-emission terms already included implicitly through the PDFs in
the leading order cross section.
In contrast, the true virtual terms are already included through the
PDFs themselves, i.e.\ through the second term in round brackets in
eq.~(\ref{eq:pdf-LO-Q0}). 

As an example, consider $pp\to Z$. At $\nLO$ we will have events such
as $gq \to Zq$, where the outgoing quark comes from collinear
initial-state splitting $g \to q\bar q$, with an underlying hard
subprocess $\bar q q \to Z$.
From these events LoopSim will generate a configuration in which the
outgoing quark is ``looped''. 
This will come in with a PDF weight that is the product of a gluon
distribution and a quark distribution, so it appears that we have a
(negative) $gq \to Z$ contribution, which would be unphysical.
However in the LO cross section with a factorisation scale $\muF \sim
Q$, when we write $\bar q q \to Z$, part of the $\bar q$ PDF comes
from $g \to \bar q q$ splitting.
If we were just to add the real $gq \to Zq$ diagram to the LO cross
section alone, then in the collinear limit we would be double counting
the part already included in the PDF. 
With the negative ``$gq \to Z$'' LoopSim contribution, what happens is
that we simply remove the $\bar q$ PDF component, generated from $g
\to \bar q q$ splitting, that was implicitly included at LO with an
incorrect final state (i.e.\ lacking an outgoing quark), since we are
now putting it in with the correct final state through the real $gq
\to Zq$ diagram.

Note that we have not yet worked out the full extension of this
discussion to higher orders. The details would depend on the precise
higher orders that we have in mind, for example \nnLO versus \nNLO.
However, regardless of these details, the fundamentally unitary nature
of the LoopSim procedure is important in ensuring that the simulated
``loops'' simply bring about an overall consistent set of final states
while maintaining the total cross section as calculated with a sensible
factorisation scale choice.

\section{Secondary emitters in LoopSim }
\label{sec:secondary-emitters}

\begin{figure}[ht]
  \centering
  \begin{psfrags}
    \psfrag{a}{(a)}
    \psfrag{b}{(b)}
    \psfrag{c}{(c)}
    \psfrag{d}{(d)}
    \psfrag{A}{$C_FC_AW_2$}
    \psfrag{B}{$C_F^2W_1$}
    \psfrag{C}{$-C_FC_AW_2$}
    \psfrag{D}{$-C_F^2W_1$}
    \psfrag{1}{$1$}
    \psfrag{2}{$2$}
    \psfrag{q}{$q$}
    \psfrag{qb}{$\bar q$}
    \includegraphics[scale=0.45]{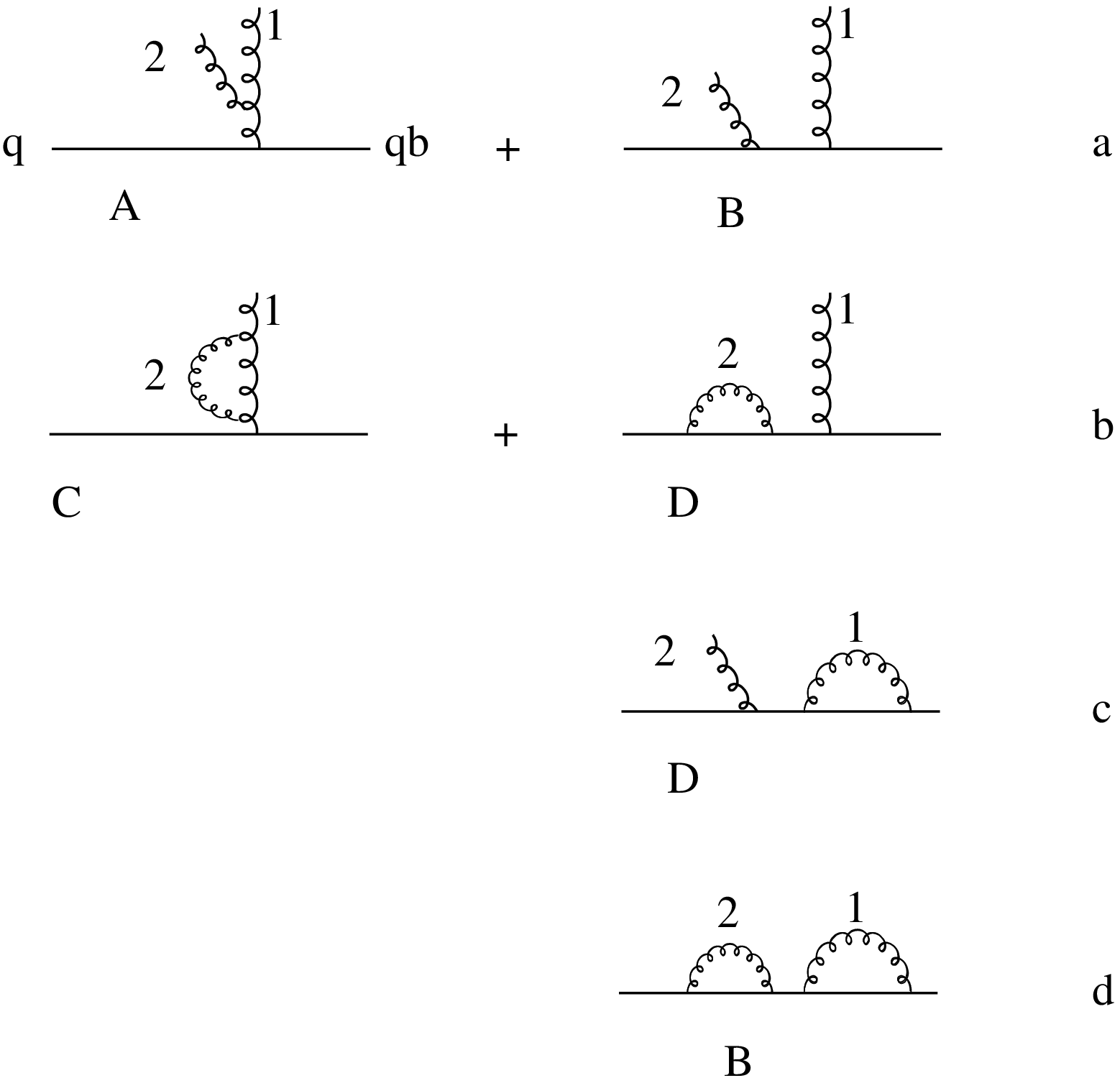}
  \end{psfrags}
  \caption{Schematic depiction of the matrix elements for two gluons
    emitted (or virtual) from a $q\bar q$ dipole: (a) gluons $1$ and
    $2$ real; (b) gluon $1$ real and $2$ virtual; (c) gluon $1$
    virtual and $2$ real; (d) gluons $1$ and $2$ virtual. In each
    case, when needed, we use the decomposition into $W_1$ and $W_2$
    pieces to separate what can be seen as the emission of
    gluon $2$ from gluon $1$, and what can be seen as the emission of
    gluon $2$ directly from $q\bar q$.}
  \label{fig:two_gluons_emission}
\end{figure}

In section~\ref{sec:looping-particles} we discussed the special
treatment needed for ``secondary emitters'', \ie non-Born particles
that have emitted something.
In our procedure, secondary emitters do not get looped:
when particle $j$ makes a loop over $i$, this is justified by the
collinear enhancement of the matrix element due to $i$ and $j$
being close in angle. But the emission of the same $j$ from the
configuration where $i$ is virtual does not have such a collinear
enhancement, so one must not take it into account.
Another way to understand it is to consider the example of
$2$-gluon emission from a $q\bar q$ dipole. The squared matrix
element for the emission of $2$ real energy-ordered ($E_1\gg E_2$)
gluons, $g_1$, $g_2$, can be expressed as
\cite{Bassetto:1984ik,Fiorani:1988by,Dokshitzer:1992ip}
\begin{equation}
  M(k_1,k_2) = (4\pi\alpha_s)^2(C_F^2W_1+C_FC_AW_2)\,,
\end{equation}
with
\begin{subequations}
  \begin{align}
    W_1 & = 4\frac{(p_q.p_{\bar{q}})}{(p_q.k_1)(k_1.p_{\bar{q}})}\frac{(p_q.p_{\bar{q}})}{(p_q.k_2)(k_2.p_{\bar{q}})}\,, \\
    W_2 & =
    2\frac{(p_q.p_{\bar{q}})}{(p_q.k_1)(k_1.p_{\bar{q}})}\left(\frac{(p_q.k_1)}{(p_q.k_2)(k_2.k_1)}+\frac{(p_{\bar{q}}.k_1)}{(p_{\bar{q}}.k_2)(k_2.k_1)}-\frac{(p_q.p_{\bar{q}})}{(p_q.k_2)(k_2.p_{\bar{q}})}\right)\,.
  \end{align}
\end{subequations}
Since the $W_1$ term diverges when $g_2$ is collinear to $q$ or $\bar
q$ (unlike the $W_2$ term), it becomes relevant when $g_2$ is
considered to have been emitted from $q$ or $\bar q$ independently of
$g_1$. The $W_2$ term diverges when $g_2$ is collinear to $g_1$
(unlike $W_1$), so it becomes relevant when $g_2$ is considered to
have been emitted from $g_1$. This is depicted in
fig.~\ref{fig:two_gluons_emission}, which also shows the virtual
corrections (cf.\ \cite{Dokshitzer:1992ip}). One notices that the
$W_2$ term only appears when $g_1$ is real. The diagrams where $g_1$
is virtual are taken into account when $g_2$ is emitted from $q$ or
$\bar q$.  Therefore, $g_1$ cannot become virtual when $g_2$ makes a
loop over it.


\end{document}